\documentclass[iop,apj,revtex4]{emulateapj} 

\newcommand{\e}[1]{\times 10^{#1}}
\newcommand{\aave}[1]{\left\langle#1\right\rangle}
\usepackage{color}
\usepackage{pslatex}

\renewcommand{\vec}[1]{\mathbf{#1}}

\newcommand{\ex}{\vec e_x}
\newcommand{\ey}{\vec e_y}
\newcommand{\ez}{\vec e_z}
\newcommand{\eb}{\vec{\hat b}}
\newcommand{\eref}[1]{(\ref{#1})}

\renewcommand{\(}{\left(}
\renewcommand{\)}{\right)}
\renewcommand{\[}{\left[}
\renewcommand{\]}{\right]}

\begin{document}

\title{Direct Numerical Simulations of Reflection-Driven, Reduced MHD
  Turbulence from the Sun to the Alfv\'en Critical Point}

\author{Jean Carlos Perez}
\author{Benjamin D. G. Chandran}
\affiliation{${~}^1$Space Science Center, University of New Hampshire, Durham, NH 03824}

\begin{abstract}
  We present direct numerical simulations of inhomogeneous reduced magnetohydrodynamic (RMHD) turbulence between the Sun and the Alfv\'en critical point. These are the first such simulations that take into account the solar-wind outflow velocity and the radial inhomogeneity of the background solar wind without approximating the nonlinear terms in the governing equations. Our simulation domain is a narrow magnetic flux tube with a square cross section centered on a radial magnetic field line. We impose periodic boundary conditions in the plane perpendicular to the background magnetic field~$\vec{B}_0$. RMHD turbulence is driven by outward-propagating Alfv\'en waves ($z^+$ fluctuations) launched from the Sun, which undergo partial non-WKB reflection to produce sunward-propagating Alfv\'en waves ($z^-$ fluctuations). Nonlinear interactions between $z^+$ and $z^-$ then cause fluctuation energy to cascade from large scales to small scales and dissipate. We present ten simulations with different values of the correlation time~$\tau_{\rm c\,\sun}^+$ and perpendicular correlation length~$L_{\perp \sun}$ of outward-propagating Alfv\'en waves (AWs) at the coronal base. We find that between 15\% and 33\% of the~$z^+$ energy launched into the corona dissipates between the coronal base and Alfv\'en critical point, which is at $r_{\rm A} = 11.1 R_{\sun}$ in our model solar wind. Between 33\% and 40\% of this input energy goes into work on the solar-wind outflow, and between 22\% and~36\% escapes as~$z^+$ fluctuations through the simulation boundary at $r=r_{\rm A}$. Except in the immediate vicinity of~$r = R_{\sun}$, the $z^\pm$ power spectra scale like~$k_\perp^{-\alpha^\pm}$, where $k_\perp$ is the wavenumber in the plane perpendicular to~$\vec{B}_0$.  In our simulation with the smallest value of $\tau_{\rm c\,\sun}^+$ ($\sim 2 \mbox{ min}$) and largest value of~$L_{\perp \sun}$ ($2\times 10^4 \mbox{ km}$), we find that $\alpha^+$ decreases approximately linearly with increasing~$\ln(r)$, reaching a value of~1.3 at $r=11.1 R_{\sun}$. Our simulations with larger values of~$\tau_{\rm c\,\sun}^+$ exhibit alignment between the contours of constant~$\phi^+$, $\phi^-$, $\Omega_0^+$, and $\Omega_0^-$, where $\phi^\pm$ are the Els\"asser potentials and $\Omega_0^\pm$ are the outer-scale parallel Els\"asser vorticities. This alignment reduces the efficiency of nonlinear interactions at $r\gtrsim 2 R_{\sun}$ to a degree that increases with increasing~$\tau_{\rm c\,\sun}^+$.
\end{abstract}

\maketitle

\vspace{0.2cm} 
\section{Introduction}
\label{sec:intro} 
\vspace{0.2cm} 

The solar wind is pervaded by turbulent fluctuations in the velocity,
density, magnetic field, and electric
field~\citep{coleman68,belcher71a,coles89,bale05,wan12}.  These
fluctuations are only weakly compressive, in the sense that the
fractional density fluctuations $\delta n/ n$ are much less
than~$|\delta \vec{B}|/B$, where $\delta \vec{B}$ is the fluctuating
magnetic field and $\vec{B}$ is the local magnetic
field~$\vec{B}$~\citep{tumarsch95}.  Within the inertial range,
$|\delta \vec{B}| \ll B$, $\delta \vec{B}$ is nearly perpendicular
to~$\vec{B}$, and $\delta \vec{B}$ varies more rapidly in the
directions perpendicular to~$\vec{B}$ than in the direction
along~$\vec{B}$~\citep{matthaeus90,tumarsch95,sahraoui09,chen12a}.
Together, these properties imply that most of the fluctuation energy
in the inertial range can be described within the framework of reduced
magnetohydrodynamics
(RMHD)~\citep{kadomtsev74,strauss76,zank92,schekochihin09}.  In this
paper, we focus on the RMHD-like component of solar-wind turbulence,
acknowledging that other types of fluctuations may also be
present. Because Alfv\'en waves (AWs) are the linear wave mode of
RMHD, we use the terms AWs, AW fluctuations and RMHD fluctuations
interchangeably. For the same reason, we use the phrases AW turbulence
and RMHD turbulence interchangeably.

Most of the AW fluctuations measured near Earth propagate away from
the Sun in the solar-wind frame \citep{roberts87}, which suggests that AWs
are launched into the solar wind from the Sun, consistent with
observations of AW-like phenomena in the solar
atmosphere~\citep{depontieu07,tomczyk07}. \cite{parker65} suggested
that the solar wind is largely heated and powered by AWs launched by
the Sun. Because linear damping of AWs is extremely weak at large
wavelengths~\citep{barnes66}, a turbulent cascade of AW energy from
large scales to small scales is needed if the AWs generated by the Sun
are to heat the solar wind efficiently. On the other hand, a turbulent
AW energy cascade requires nonlinear interactions between
counter-propagating AWs \citep{iroshnikov63,kraichnan65}, and the Sun
launches only outward-propagating AWs. In order for AWs to heat
coronal holes and the solar wind, some source of inward-propagating
AWs is thus needed.
\begin{figure*}[htb]
  \centering
   \includegraphics[width=\textwidth]{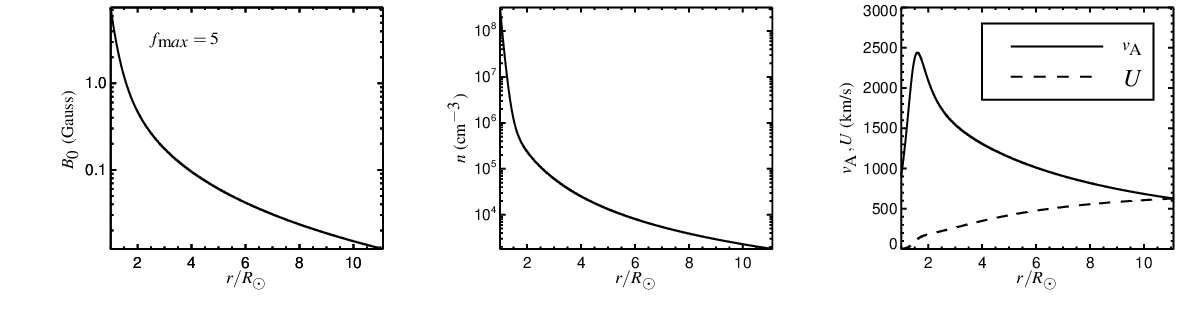}
  \caption{Radial profiles of $B_0$, $n$, $U$, and $v_{\rm A}$ used in
    our simulations.  
}\label{fig:profiles}
\end{figure*}

One such source is the non-WKB reflection of AWs due to radial
variations in the Alfv\'en speed~$v_{\rm
  A}$~\citep{heinemann80,velli93,hollweg07}. Photospheric motions
generate AWs with periods~$P$ of minutes to hours, corresponding to
parallel wavelengths (measured in the direction of the background
magnetic field~$\vec{B}_0$) of $\sim Pv_{\rm A}$, which is $\gtrsim 1
R_{\sun}$ in coronal holes~\citep{cranmer05}. These parallel wavelengths
are comparable to the gradient lengthscales of the mass density~$\rho$
and magnetic field~$\vec{B}$ in coronal holes and the near-Sun solar
wind, which causes outward-propagating AWs to undergo non-WKB reflection.

Recent work on reflection-driven RMHD turbulence falls into two
categories. The first category consists of studies that employ
simplified models of the nonlinear terms in the governing
equations~\citep{velli89,dmitruk02,cranmer05,verdini07,chandran09c}. The
most sophisticated of these are the works of \cite{verdini09a} and
\cite{verdini12}, who simulated reflection-driven RMHD turbulence from
$r=1 R_{\sun}$ out to~$r \sim 20 R_{\sun}$ using a shell model to
approximate the nonlinear terms. The second category consists of
direct numerical simulations of reflection-driven RMHD turbulence that
retain the full nonlinear terms in the equations of inhomogeneous
RMHD.  As far as we are aware, there are only two previous studies in this
category. \cite{dmitruk03} carried out simulations extending from the
coronal base to a heliocentric distance~$r$ of $2R_{\sun}$, neglecting
the solar-wind outflow velocity.  Van Ballegooijen et
al.~(2011)\nocite{vanballegooijen11} also neglected the solar-wind
outflow velocity and carried out numerical simulations extending from
the photosphere, through the chromosphere, and into the corona,
focusing on closed coronal loops.

In this work, we present the first direct numerical simulations of
inhomogeneous RMHD turbulence in coronal holes and the near-Sun solar
wind that incorporate solar-wind outflow and wave reflection without
approximating the nonlinear terms in the governing equations.  Our
simulations extend from the coronal base out to the Alfv\'en critical
point, the point at which the solar-wind outflow speed equals the
Alfv\'en speed, which is at $r=11.1 R_{\sun}$ in our model solar
wind. We describe our mathematical model in Section~\ref{sec:model}
and our numerical method in Section~\ref{sec:method}. In
Section~\ref{sec:simulations}, we analyze results from ten different
numerical simulations.  In Section~\ref{sec:anomalous} we discuss the
inertial-range power spectra in the simulations, and the possible
connection to the $\sim k^{-1}$ magnetic-field spectrum observed at
small wavenumber~$k$ in the interplanetary medium. In
Section~\ref{sec:vpalign} we describe an alignment effect that reduces
the efficiency of nonlinear interactions in several of our
simulations, and in Section~\ref{sec:conclusion} we summarize our
results and conclusions.

\vspace{0.2cm} 
\section{Open-Field RMHD model}
\label{sec:model}
\vspace{0.2cm} 

We consider RMHD turbulence in a narrow magnetic flux tube centered on
a radial magnetic field line extending outwards from the Sun. We
define~$x$ and $y$ to be Cartesian coordinates in the plane
perpendicular to this central radial magnetic field line. The essence
of our ``narrow-flux-tube approximation'' is the assumption that
\begin{equation}
\sqrt{x^2 + y^2} \ll r.
\label{eq:nft} 
\end{equation} 
That is, we restrict our analysis to the vicinity of the central
radial field line.  The equations that we present in this section can
be viewed as the leading-order terms in an expansion in powers of
$\theta_{\rm max}$, where $\theta_{\rm max}$ is an upper bound on the
spherical polar angle~$\theta$ in a spherical coordinate system in
which $\theta=0$ corresponds to the central radial magnetic field line.

We take the mass density~$\rho$, background magnetic field
strength~$B_0$, and solar-wind (proton) outflow velocity to be fixed
functions of the distance~$s$ measured along the magnetic
field. Because of Equation~(\ref{eq:nft}), $s$ is approximately equal to
the heliocentric distance~$r$, and we can write
\begin{eqnarray}
  \rho = \rho(r),\quad\vec{U} &=& U(r)\eb,\quad \vec{B}_0=B_0(r)\eb,
  \label{eq:radial}
\end{eqnarray}
where $\eb$ is the unit vector of the background magnetic field. The
middle equality in Equation~(\ref{eq:radial}) expresses our assumption
that~$\vec{U}$ is parallel to~$\vec{B}_0$.  Since the flux tube is
narrow, $\eb$ is approximately radial, and also approximately parallel
to the magnetic field line at $x=y=0$.  We adopt the same solar-wind
profile as \cite{chandran09c}, setting $\rho = m_{\rm p} n$ and
\begin{equation}
n(r) = \left(\frac{3.23 \times 10^8}{\tilde{r}\,^{15.6}} + \frac{2.51
  \times 10^6}{\tilde{r}\,^{3.76}} + \frac{1.85 \times
  10^5}{\tilde{r}\,^2}\right) \mbox{ cm}^{-3},
\label{eq:n} 
\end{equation} 
where $\tilde{r} = r/R_{\sun}$, $m_{\rm p}$ is the proton mass,
and~$n$ is the proton number density. The first two terms on the
right-hand side of Equation~(\ref{eq:n}) are from \cite{feldman97},
and the $r^{-2}$ term has been added so that $n$ extrapolates to the
value $4 \mbox{ cm}^{-3}$ at $r=1 \mbox{ AU}$.  The background field
strength is given by
\begin{equation}
B_0 = \left[\frac{1.5(f_{\rm max} - 1)}{\tilde{r}\,^6} +
  \frac{1.5}{\tilde{r}\,^2}\right] \mbox{ Gauss},
\label{eq:B0} 
\end{equation} 
where $f_{\rm max}$ is the usual super-radial expansion factor, which
we set equal to~5. The solar-wind speed then follows from flux
conservation,
\begin{equation}
U = 9.25 \times 10^{12} \;\frac{B_{\rm G}}{\tilde n}
\;\mbox{cm}\;\mbox{s}^{-1} ,
\label{eq:defU} 
\end{equation} 
where $B_{\rm G}$ is $B_0$ in Gauss and $\tilde n$ is $n$ in units of
$\mbox{cm}^{-3}$.  The normalization constant in
Equation~(\ref{eq:defU}) has been chosen so that $U = 750\;
\mbox{km}\; \mbox{s}^{-1}$ at 1~AU.  The Alfv\'en critical point in
this model is at $r_a=11.1 R_{\sun}$, and the maximum of $v_{\rm A}$
is at $r_m= 1.60 R_{\sun}$. The magnetic field strength, density,
solar-wind speed, and Alfv\'en speed are plotted in
Figure~\ref{fig:profiles}.

We consider non-compressive fluctuations in the velocity~$\delta
\vec{v}$ and magnetic field~$\delta \vec{B}$ within this narrow flux
tube. We assume that $\delta \vec{v} \cdot \eb = 0$, $\delta \vec{B} \cdot
\eb = 0$ and that the fluctuations vary on lengthscales in the
directions perpendicular to~$\vec{B}_0$ that are much smaller than~$r$
and much smaller than the lengthscales on which the fluctuations vary
in the direction parallel to~$\vec{B}_0$. We further assume that
$\delta B \ll B_0$ and that the characteristic lengthscales and
timescales of the fluctuations are $\gg \rho_{\rm p}$ and $\Omega_{\rm
  p}^{-1}$, where $\rho_{\rm p}$ and $\Omega_{\rm p}$ are the proton
gyroradius and cyclotron frequency.  Given these assumptions, the
fluctuations are well described by the equations of inhomogeneous
RMHD~\citep{heinemann80,velli89}, which can be written in the form 
\begin{eqnarray}
  \frac{\partial\vec{z}^\pm}{\partial t}+\left(U\pm
  v_{\rm A}\right)\frac{\partial \vec{z}^\pm}{\partial r}+\left(U\mp
  v_{\rm A}\right)\left(\frac {\vec{z}^\pm}{4H_\rho}-\frac
  {\vec{z}^\mp}{2H_{\rm A}}\right)\nonumber\\
 =-   \vec{z}^\mp\cdot\vec{\nabla}_\perp\vec{z}^\pm - \frac{\nabla_\perp P}\rho -\nu_p (-\nabla_\perp^{2})^p \vec{z}^\pm,
\label{eq:velli2} 
\end{eqnarray} 
 where
\begin{eqnarray}
  \vec{z}^\pm &=& \delta \vec{v} \mp \frac{\delta \vec{B}}{\sqrt{4\pi \rho}},\\
  \label{eq:zpm} 
  \nabla_\perp &=& \ex \frac{\partial}{\partial x} + \ey \frac{\partial}{\partial y},\\
\label{eq:defnablaperp} 
H_\rho^{-1} &=& - \rho^{-1} d\rho/dr, \\
H_B^{-1} &=& - B_0^{-1} dB_0/dr,\\
H_{\rm A}^{-1} &=& v_{\rm A}^{-1} dv_{\rm A}/dr,
\end{eqnarray}
$p$ is an integer $\geq 1$, and $\nu_p(r)$ is a positive definite hyperviscosity
coefficient. The hyperviscosity term in Equation~(\ref{eq:velli2}),
which will be discussed in more detail in Section~\ref{sec:method},
has been added to act as a sink of energy in the system at small
(perpendicular) lengthscales to ensure numerical stability.

Equation~(\ref{eq:velli2}) contains a key piece of physics that is not
present in the homogeneous RMHD model, namely the linear coupling
between~$\vec z^+$ and $\vec z^-$ fluctuations through the $H_{\rm
  A}^{-1}$ term. This linear coupling is responsible for the non-WKB
reflection of AWs. As $H_{\rho}$ and $H_{\rm A}$ are increased to
infinity, Equation~(\ref{eq:velli2}) reduces to the usual equations of
homogeneous~RMHD.

The Els\"asser potentials $\phi^\pm$ are related to~$\vec{z}^\pm$
through the equation
\begin{equation}
\vec z^\pm=\eb\times\nabla\phi^\pm\label{eq:potential},
\end{equation}
and the field-aligned Els\"asser vorticity is given by
\begin{equation}
\Omega^\pm=\eb\cdot\nabla\times\vec z^\pm=\nabla_\perp^2\phi^\pm.
\label{eq:vorticity-potential}
\end{equation}
By taking the curl of \eref{eq:velli2} and taking the dot
product of the resulting equation with $\eb$, we obtain
\begin{eqnarray}
  \frac{\partial\Omega^\pm}{\partial t}&+&\left(U\pm
  v_{\rm A}\right)\(\frac{\partial\Omega^\pm}{\partial s}-\frac{\Omega^\pm}{2H_B}\)\nonumber\\
  &+&\left(U\mp
  v_{\rm A}\right)\left(\frac 1{4H_\rho}\Omega^\pm-\frac
  1{2H_{\rm A}}\Omega^\mp\right)\nonumber\\
 &=&-\eb\cdot\left[\nabla\times(\vec{z}^\mp\cdot\vec{\nabla}\vec{z}^\pm)\right]-\nu_p(-\nabla_\perp^2)^{p}\Omega^\pm\nonumber.\\
\label{eq:vorticity} 
\end{eqnarray}
Equation~\eref{eq:vorticity} extends the RMHD model for coronal loops
of \cite{vanballegooijen11} to include a background solar wind
flow~$U$.

\vspace{0.2cm} 
\section{Numerical Method and the IRMHD Code}
\label{sec:method} 
\vspace{0.2cm} 

\begin{figure}[!t]
\begin{center}
\vspace*{-0.1in}
\includegraphics[width=0.45\textwidth]{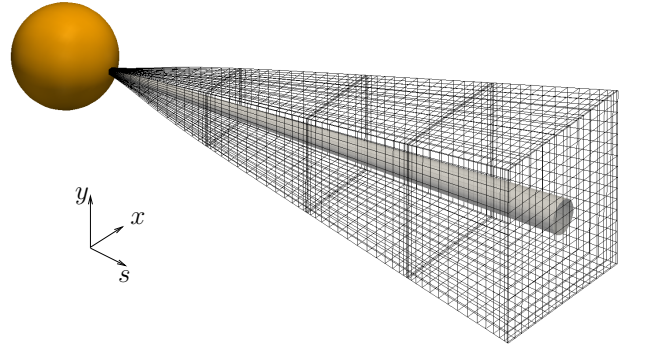}
\end{center}
\caption{\footnotesize Numerical domain and coordinates
  employed by our IRMHD turbulence code. (See text for details.) 
\label{fig:sim_dom} }
\end{figure}

We solve Equation~(\ref{eq:vorticity}) in a simulation domain that
consists of a narrow magnetic flux tube with a square cross section
extending from a heliocentric distance of $r_{\rm min} = 1 R_{\sun}$
(the coronal base) out to the Alfv\'en critical point $r_{\rm max} =
r_{\rm A}$ ($11.1 R_{\sun}$ in our model) at which $U= v_{\rm A}$, as
illustrated in Figure~\ref{fig:sim_dom}.  We define $L_\perp(r)$ to be
the dimension of the simulation domain perpendicular to~$\vec{B}_0$ at
heliocentric distance~$r$, and $L_{\perp \sun}$ to be the value of
$L_\perp(r)$ at $r=r_{\rm min}$.  Because the magnetic flux through
the flux-tube cross section is independent of~$r$,
\begin{equation} 
 L_\perp(r) = L_{\perp \sun}\sqrt{\frac{B_0(r_{\rm min})}{B_0(r)}}.
  \label{eq:Rexpansion}
\end{equation} 
Given Equation~(\ref{eq:B0}), $L_\perp$ expands by a factor of 24.8
between~$r_{\rm min}$ and $r_{\rm max}$.  In half of our simulations,
we set $L_{\perp \sun} = 10^4$~km, and in the other half  we
set~$L_{\perp \sun} = 2\times 10^4$~km.  At all~$r$, $L_{\perp}/r <
0.032$ when $L_{\perp \sun} = 10^4 \mbox{ km}$, and $L_{\perp}/r < 0.064$
when $L_{\perp \sun} = 2\times 10^4 \mbox{ km}$, consistent with 
Equation~(\ref{eq:nft}).

Because $L_\perp \ll r$, we can neglect the curvature of surfaces
perpendicular to~$\vec{B}_0$ when solving
Equation~(\ref{eq:vorticity}) and treat these surfaces as planes. At
each~$r$, we employ $N_\perp$ uniformly spaced grid points in each of
the $x$ and $y$ directions, where $N_\perp$ is independent of~$r$.
Since $L_\perp(r)$ increases with~$r$, the grid spacing in $x$ and~$y$
increases with~$r$.  We take the $\vec{z}^\pm$ fluctuations to satisfy
periodic boundary conditions in the $x-y$ plane. However, as we will
discuss further below, radial inhomogeneity prevents us from using
periodic boundary conditions in the~$r$ direction.

Equation~(\ref{eq:vorticity}) represents a system of two coupled
partial differential equations (PDEs), one for $\Omega^+(\vec
x_\perp,r,t)$ and one for~$\Omega^-(\vec x_\perp,r,t)$, where $\vec
x_\perp = x \ex + y \ey$ is the position vector in the $x-y$~plane.
The code that we have developed for this investigation, called the
Inhomogeneous RMHD Code, or IRMHD Code, solves this system of PDEs
using a spectral element method (SEM) based on a Chebyshev-Fourier
basis~\citep{CanHus88}.  The essence of the SEM method is to perform a
decomposition of the domain along the radial direction into $M$
subdomains $\mathcal D_n$, where $n=1,2,\ldots,M$. The Els\"asser
vorticities (and, analogously, $\vec{z}^\pm$ and $\phi^\pm$) are
approximated by a truncated Chebyshev-Fourier expansion in each
subdomain,
\begin{equation}
  \left.\Omega^\pm(\vec{x}_\perp, r,t)\right|_{\mathcal D_n} =
  \sum_{\alpha\tilde{k}_x \tilde{k}_y} \Omega^\pm_{\alpha \tilde{k}_x \tilde{k}_y,n}(t)T_\alpha(\xi_n) \exp(i\mathbf{k_\perp} \cdot
  \vec{x}_\perp),
\label{eq:expansion} 
\end{equation}
where 
\begin{equation}
\vec{k}_\perp = \frac{2 \pi \tilde{\vec{k}}_\perp }{L_{\perp}(r)}
\label{eq:defkperpt} 
\end{equation} 
is the perpendicular wavevector, $\tilde{\vec{k}}_\perp = \tilde{k}_x
\ex + \tilde{k}_y \ey$, and $T_\alpha(\xi_n) = \cos (\alpha
\cos^{-1}\xi_n)$ is the Chebyshev polynomial of order~$\alpha$. The
quantity $\xi_n = 2(r - r_n)/\Delta r_n$ is the normalized radial
coordinate in sub-domain $\mathcal D_n$, where $r_{n}$ is the radial
midpoint of the $n^{\rm th}$ subdomain, and $\Delta r_n$ is the length
of the $n^{\rm th}$ subdomain in the $r$~direction.  Within the
$n^{\rm th}$ subdomain, we discretize the radial interval using the
Gauss-Lobatto grid, $\xi_{n,j}= \cos(\pi j/N_d)$, where $N_d$ is the number of radial grid points per subdomain.  This choice of grid enables us to
carry out the Chebyshev transform using a fast cosine transform and to
retain the exponential accuracy characteristic of spectral methods.
The quantities $\tilde{k}_x$, $\tilde{k}_y$, and~$\alpha$ take on
integer values only, with $-N_\perp/2+1 \leq \tilde{k}_x \leq N_\perp
/2$, $-N_\perp/2+1\leq \tilde{k}_y \leq N_\perp /2$, and $0 \leq \alpha
< N_d$. This Chebyshev-Fourier expansion results in a
system of ordinary differential equations (ODEs) for the
Chebyshev-Fourier coefficients $\Omega^\pm_{\alpha \tilde{k}_x
  \tilde{k}_y,n}(t)$ in each subdomain.  These equations are coupled
through boundary conditions (continuity of $\vec{z}^\pm$) at the
interface surfaces. 

As discussed in Section~\ref{sec:model}, the hyperviscosity term 
in Equation~\eref{eq:velli2}  dissipates the fluctuations 
before they can cascade to the grid-scale. We set
\begin{equation}
  \nu_p(r) = \tilde\nu_p\left[\frac{L_\perp(r)}{2\pi}\right]^{2p}
\label{eq:nup} 
\end{equation}
with $p=4$ and $\tilde\nu=5\e{-5}z^+_{\rm{rms}\sun}/L_\sun$, where
$z^+_{\rm{rms}\sun}$ is the imposed amplitude of $z^+$ fluctuations at
$r=R_\sun$ (see section~\ref{sec:simulations}). The Chebyshev-Fourier
coefficients of the hyperviscosity term in
Equation~(\ref{eq:vorticity}) within the $n^{\rm th}$ radial subdomain
are then simply $-\tilde \nu_p \tilde{k}_\perp^{2p} \Omega^\pm_{\alpha
  \tilde{k}_x \tilde{k}_y,n}$. We advance the solution to
Equation~(\ref{eq:vorticity}) forward in time using a third-order
Runge-Kutta method and employ an integrating factor to handle the
hyperviscosity term. With this approach, the time step is not limited
by the hyperviscous timescale, and is instead constrained solely by
the accuracy and stability requirements of the non-dissipative
terms. For initial conditions, we set
\begin{equation}
  \left.\Omega^\pm(\vec{x}_\perp, r,t=0)\right|_{\mathcal D_n} = 0
\label{eq:IC} 
\end{equation}
at all $\vec{x}_\perp$ and~$r$.

The IRMHD code uses the Message Passing Interface (MPI) programming
paradigm and possesses excellent scaling properties on massively
parallel supercomputers due to the nature of the domain decomposition
in the SEM algorithm.  To parallelize the code, we assign each
subdomain to a different set of processors, with the appropriate
communications between domains to transfer boundary information,
maintaining a low network overhead. The single-domain component of the
IRMHD code is a fully de-aliased 3D Chebyshev-Fourier pseudo-spectral
algorithm that performs spatial discretization on a grid with
$N_\perp^2\times N_d$ grid points. This decomposition
results is a global mesh of $N_\perp^2\times N_r$ grid points where
$N_r=(N_d-1)M+1$ is total number of radial points from all the
subdomains\footnote{More precisely, the Chebyshev transform involves
both the inner and outer boundary, which overlap for contiguous
subdomains. Therefore the full radial domain consists of $N_dM$ radial
grid points minus $(M-1)$ overlapping boundaries.}. For the
simulations described in Section~\ref{sec:method}, $N_d = 17$ and $M =
512$, which leads to $N_r = 8193$, and $\Delta r_n$ is taken to be the
same for each subdomain, which leads to~$\Delta r_n \simeq 0.02
R_{\sun}$.  Within each subdomain, we parallelize the 2D fast Fourier
transform in the $x-y$ plane. In low-resolution runs, we turn off this
``intra-domain'' parallelization, assigning one processor per
subdomain.  However, for the runs reported in this paper, 16
processors per subdomain were used, requiring a total of 8192
processors in a single simulation.  We have verified the accuracy of
the IRMHD code through extensive tests of linear wave propagation and
reflection and nonlinear conservation of wave~action.

\subsection{Radial Boundary conditions}

The equations describing the evolution of $\Omega^+$ and $\Omega^-$
are hyperbolic advection equations, where $\Omega^+$ is advected at
radial velocity $U+v_{\rm A}$ and $\Omega^-$ is advected at radial
velocity~$U-v_{\rm A}$.  The radial velocity of~$\Omega^-$ is negative
at $r<r_{\rm A}$, where $U< v_{\rm A}$, and positive at $r>r_{\rm
  A}$. Given that $U+v_{\rm A}>0$, one must specify one boundary
condition on~$\Omega^+$ at the lower boundary $r= R_{\sun}$ to solve
the advection PDE for~$\Omega^+$. This boundary condition determines
the properties of the $\vec z^+$ fluctuations that are advected into
the simulation domain at $r=R_{\sun}$ and contains all our assumptions
about the properties of the Alfv\'en waves that are launched by the
Sun, including their amplitudes, perpendicular wavelengths, and
frequencies. It would be natural to expect that an additional boundary
condition is required to solve the advection equation for $\Omega^-$
at the outer boundary of the simulation at $r=r_{\rm max}$ to
determine the amplitudes of the inward waves that are injected towards
the Sun from radii exceeding~$r_{\rm max}$. However, an outer boundary
condition on~$\vec z^-$ is only necessary if $r_{\rm max} < r_{\rm
  A}$, in which case the radial velocity $U-v_{\rm A}$ of the $\vec
z^-$ waves is negative at~$r_{\rm max}$.  In the simulations that we
present in this work, we set $r_{\rm max} = r_{\rm A}$, and thus $\vec
z^-$ waves do not flow into the simulation domain at $r=r_{\rm
  A}$. Mathematically, this means that no additional boundary
condition needs to be (or in fact can be) imposed on $\vec z^-$ at the
outer boundary. If instead we were to set $r_{\rm max} < r_{\rm A}$
and impose an extra outer boundary condition on $\vec z^-$, then this
boundary condition would amount to an unphysical assumption that would
modify the correct solution that arises when the Alfv\'en critical
point is included in the domain.

At $r=r_{\rm min}$, we impose the random, time-dependent boundary condition
 $\vec z^+(\vec x,r=r_{\rm min},t) =
\vec z_{\rm b}^+(\vec x_\perp,t)$, where
\begin{equation} 
  \vec z_{\rm b}^+(\vec x_\perp,t)  = \sum_{\tilde{k}_x \tilde{k}_y}i\,\eb\times\vec k_\perp\phi_{b\tilde{k}_x \tilde{k}_y}(t)\exp{\(i\vec k_\perp\cdot\vec x_\perp\)}\label{eq:zbase},
\end{equation} 
\begin{equation}
  \phi_{b \tilde{k}_x \tilde{k}_y}=\left\{\begin{array}{cc}
      C\zeta_{\tilde{k}_x \tilde{k}_y}(t) & 1\le \tilde{k}_\perp\le 3\\
&\\
      0                     & \hbox{otherwise}
  \end{array}\right.\label{eq:potential-boundary}.
\end{equation}
 We set\footnote{This
  choice corresponds to a $k_\perp^{-3/2}$ dependence for the 1D
  energy spectrum, which will be defined later in
  Section~\ref{sec:simulations}} $\zeta_{\tilde{k}_x\tilde{k}_y}(t_j) =
\tilde k_\perp^{-9/4}\exp{\(i\psi_{j,\tilde{k}_x \tilde{k}_y}\)}$, where $t_{\rm j} =
j \tau_{\rm b}$, $j = -1, 0, 1, 2, \dots$, and $\psi_{j,\tilde{k}_x \tilde{k}_y}$ are
uniformly distributed random phases between $0$ and~$2\pi$.  At values
of~$t$ between consecutive~$t_{\rm j}$, we
determine~$\zeta_{\tilde{k}_x \tilde{k}_y}(t)$ using the
cubic-interpolation algorithm described
by~\cite{keys81}.  
We adjust the constants~$C$ and~$\tau_{\rm b}$ to
control the rms value of $z^+$ at the inner boundary, denoted
$z^+_{\rm rms\,\sun}$, and the correlation time~$\tau_{\rm c\,\sun}^+$
of $z^+$ at $r=R_{\sun}$ (defined in Equation~(\ref{eq:deftauc})
below).  As a rough rule of thumb, $\tau_{\rm c\,\sun}^+ \simeq 0.5
\tau_{\rm b}$. Because $z^+=0$ at all points inside the domain at
$t\leq 0$, we choose the coefficients $\zeta_{\tilde{k}_x
  \tilde{k}_y}(t_j)$ for $j=-1,0$ and $1$ in such a way that both
$\vec z_{\rm b}^+(\vec x_\perp,t=0)$ and $\partial_t\vec z_{\rm
  b}^+(\vec x_\perp,t=0)$ vanish. This prevents the propagation of
abrupt radial variations into the domain at early times.

\begin{table}[!t]
\begin{center}
  \begin{tabular}{ccccccc}
    Simulation & $L_{\perp \sun}$ & $\tau^+_{\rm c \,\sun}$ &
    $z^+_{\rm rms \, \sun}$ & $\tau_{nl,\sun}^+$&$\tau_{nl,\sun}^-$ & Run time\vspace{0.05cm}   \\
 & ($10^3$ km) & (min) & (km/s) & (min) &
    (min) & (h)\vspace{0.1cm}  \\
    \hline\hline \\
  & & & & \vspace{-0.6cm} \\
    A1    &  10     &   19.7      &42&3.01 &0.66 & 16.3       \\
    A2    &  20     &   22.4      &40&4.85 &1.32 & 16.4         \\
    B1    &  10     &   9.9       &42&2.78 &0.66 & 11.8         \\
    B2    &  20     &   9.9       &40&4.35 &1.32 & 13.3          \\
    C1    &  10     &   6.0       &42&2.59 &0.66 & 11.5          \\
    C2    &  20     &   6.7       &40&4.03 &1.32 & 12.1          \\
    D1    &  10     &   3.3       &41&2.33 &0.66 & 11.6          \\
    D2    &  20     &   3.3       &40&4.18 &1.33 & 12.9          \\
    E1    &  10     &  2.0        &41&2.40 &0.67 & 33.4          \\
    E2    &  20     &  2.0        &40&4.34 &1.34 & 31.2\vspace{0.1cm}       \\
    \hline
  \end{tabular}
\end{center}
\caption{Simulation labels and parameter values. The quantity
  $L_{\perp\sun}$ is the width of the simulation-domain cross section
  at $r = R_{\sun}$, $\tau^+_{\rm c\,\sun}$ is the the correlation
  time of the $z^+$ AWs launched into the simulation domain at $r =
  R_{\sun}$, $\tau_{\rm nl,\sun}^\pm = (k_{\perp\sun}z^\mp_{\rm rms})^{-1}$ is the
  nonlinear timescale of $z^\pm$ fluctuation at $r= R_{\sun}$, and the
  run time is $t_{\rm f} - t_{\rm i}$, where $t_{\rm i} =0$ and
  $t_{\rm f}$ are the values of $t$ at the beginning and end of each
  simulation. 
 \label{tab:simlist}}
\end{table}

As mentioned previously, since we choose $r_{\rm max}\ge r_{\rm A}$ no
additional boundary condition  needs to be imposed for $\vec z^-$. 
The value of $z^-$ at the the
Alfv\'en critical point, which we denote $z^-_{\rm A}$, satisfies the equation
\begin{equation}
  \frac{\partial\vec z_{\rm A}^-}{\partial t}+2v_{\rm A}\left(\frac
    1{4H_\rho}\vec z_{\rm A}^--\frac 1{2H_{v_{\rm A}}}\vec z_{\rm
      A}^+\right) =-\left(\vec 
    z_{\rm A}^+\cdot\nabla\right)\vec z_{\rm A}^-,
\label{eq:zmrA} 
\end{equation}
where $\vec z_{\rm A}^+$ is the value of $\vec z^+$ at $r=r_{\rm A}$.
At $t=0$, $z_{\rm A}^-=0$, from Equation~(\ref{eq:IC}).  To determine
$\vec z_{\rm A}^-$ at all subsequent times, we integrate
Equation~(\ref{eq:zmrA}) forward in time, using the time-evolving
numerical solution for $\vec z^+$ at $r=r_{\rm A}$. Note that the pressure and dissipation terms have been omitted for
simplicity

\begin{figure*}[!t]
  \centering
  \begin{tabular}{cc}
    \includegraphics[width=0.48\textwidth]{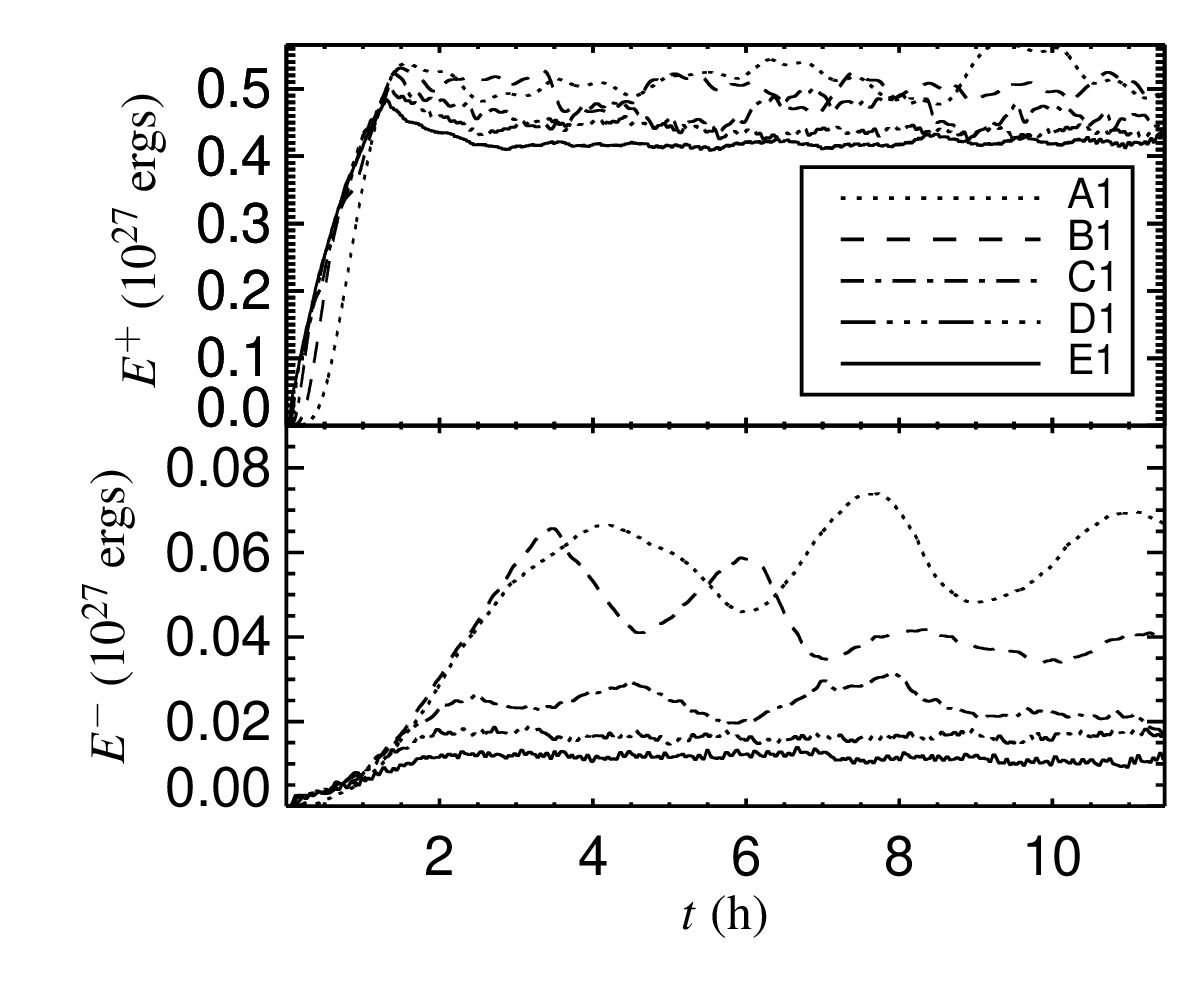}&\includegraphics[width=0.48\textwidth]{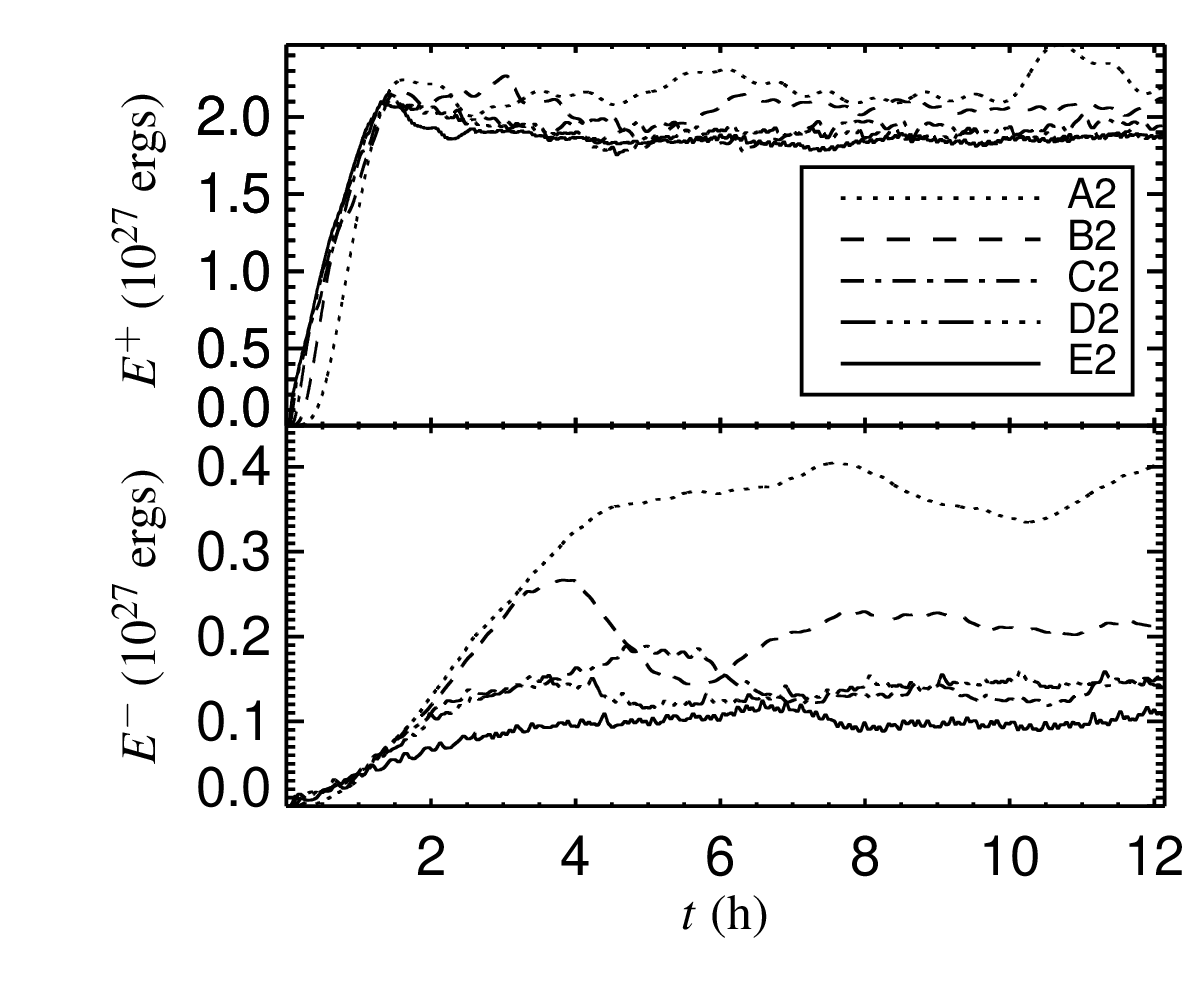}    
  \end{tabular}
  \caption{Time evolution of $E_{\rm tot}^\pm$ for the simulations in
    Table~\ref{tab:simlist}.   
\label{fig:Esaturation} }
\end{figure*}

\vspace{0.2cm} 
\section{Numerical Simulations}
\label{sec:simulations} 
\vspace{0.2cm}

In this section, we report the results of ten numerical simulations
carried out using the numerical method described in
Section~\ref{sec:method}. Each simulation uses $256^2 \times 8193$
grid points.  There are three adjustable physical parameters in the
simulations: the rms amplitude and correlation time, $z^+_{\rm
  rms\,\sun}$ and ~$\tau_{\rm c\,\sun}$, of the $z^+$ waves that are
injected into the simulation domain at $r=R_{\sun}$ and the
width~$L_{\perp \sun}$ of the cross section of the simulation domain
at~$r=R_{\sun}$.  The values of these parameters in each simulation
are listed in Table~\ref{tab:simlist}.  The values chosen for
$\tau_{\rm c \sun}^+$ are motivated by observations of magnetic bright
points and photospheric motions, which suggest that the dominant
timescales of AWs launched into the Sun range from minutes to
hours~\citep{cranmer05,verdini07}.  Previous studies of AW launching
by the Sun have considered values for the dominant AW correlation
length perpendicular to~$\vec B_0$ at $r=r_{\rm min}$ ranging from
granular scales~$\sim 10^3 \mbox{ km}$ \citep{cranmer05, hollweg10} to
scales~$\sim 5\times 10^3 \mbox{ km}$ comparable to the spacing of
photospheric flux tubes~\citep{chandran09c} to scales of $10^4 \mbox{
  km} - 3\times 10^4 \mbox{ km}$ corresponding to the diameters of
supergranules~\citep{dmitruk02,verdini07,verdini12}.  In our
simulations, this perpendicular correlation length is~$\sim
L_{\perp\sun}$, and we set $L_{\perp\sun } = 10^4$~km or $L_{\perp
  \sun} = 2\times 10^4 \mbox{ km}$.  For all of the simulations
reported in this paper, we take $z^+_{\rm b, rms} = 40 \mbox{ km/s}$.
This choice results in an rms velocity fluctuation at $r=r_{\rm min}$,
denoted $\delta v_{\rm rms \sun}$, that satisfies
\begin{equation}
\delta v_{\rm rms\,\sun} \simeq 20 \mbox{ km/s},
\label{eq:dvb} 
\end{equation} 
in agreement with {\em Hinode} measurements of the transverse motions of
field-aligned structures in the low corona~\citep{depontieu07}.

Although the values we have used for $\tau_{\rm c\,\sun}$, $L_{\perp
  \sun}$, and $z^+_{\rm rms \sun}$ are plausible, there is
considerable uncertainty in the values of these parameters in the
solar atmosphere. For example, the moving spicules observed by
\cite{depontieu07} have proton densities $\gtrsim 10^{13} \mbox{
  cm}^{-3}$ that are much larger than the value~$n_{\rm p} \simeq 3
\times 10^8 \mbox{ cm}^{-3}$ that typifies coronal holes at $r\simeq 1
R_{\sun}$~\citep{feldman97}.  The mass loading of these spicules may
slow their transverse motions relative to the motions of the
surrounding lower-density regions.  In addition, the Sun likely
launches AWs with a broad spectrum of timescales and perpendicular
correlation lengths, a point to which we return in
Section~\ref{sec:anomalous}. A more comprehensive exploration of
wave-launching scenarios would be useful, but is beyond the scope of
this paper.

\begin{figure*}[t]
  \centering
  \includegraphics[width=0.475\textwidth]{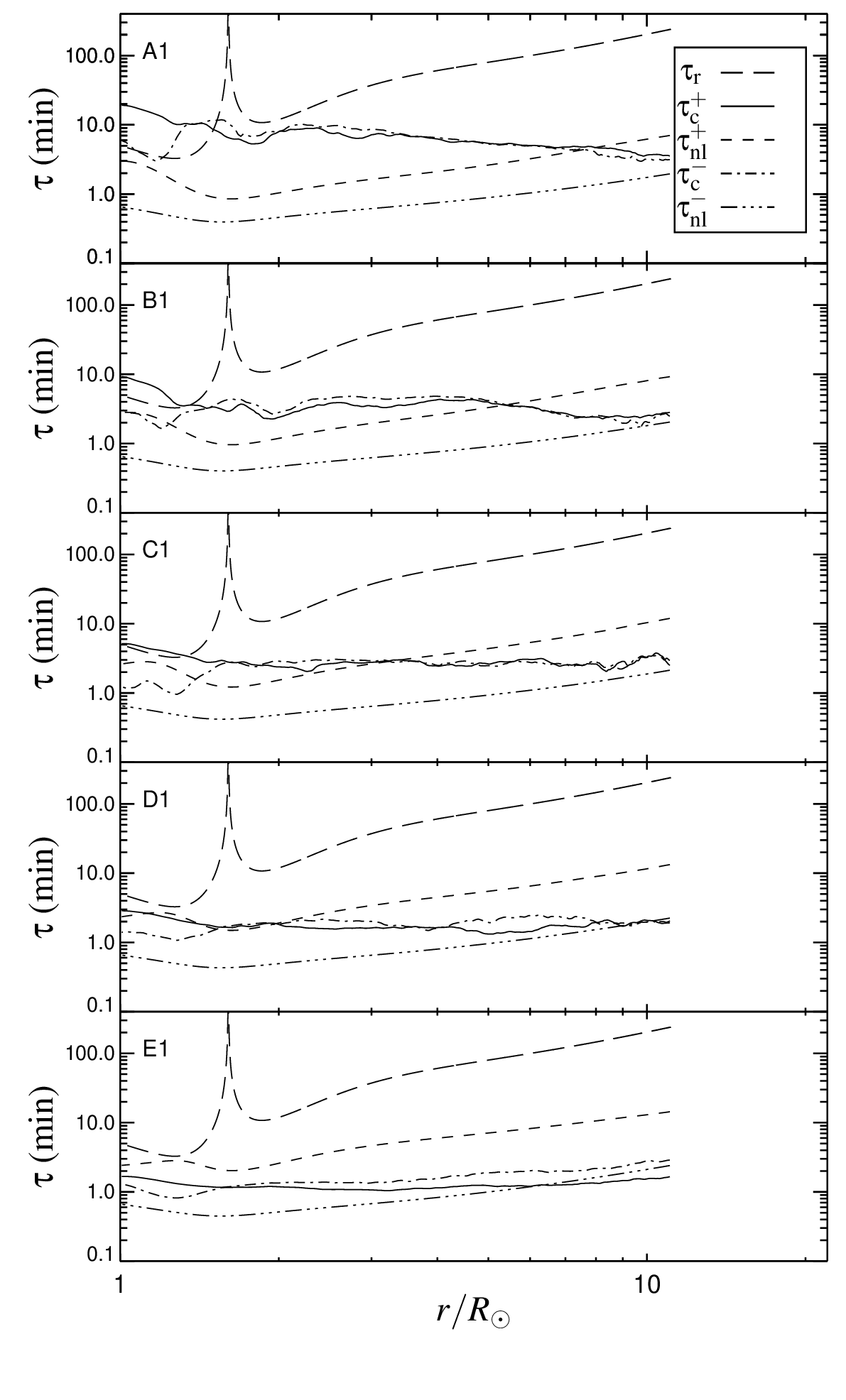}
  \includegraphics[width=0.475\textwidth]{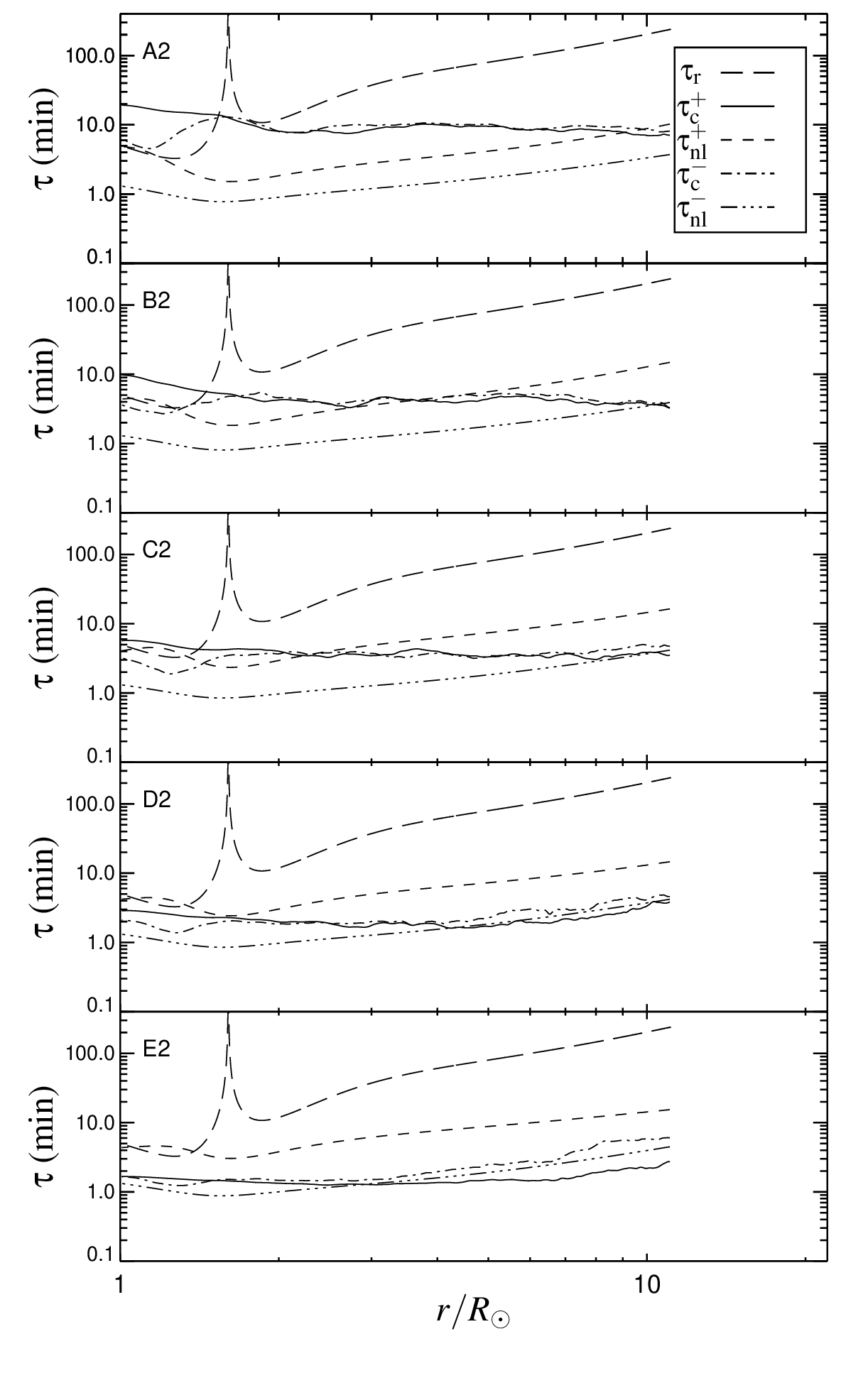}
  \caption{Radial dependence of relevant timescales in the
    simulations: The reflection timescale $\tau_r$ (long-dash),
    $z^+$ autocorrelation time $\tau^+_{\rm c}$ (solid), $z^-$
    autocorrelation time $\tau^-_{\rm c}$ (dash-dot), the nominal eddy
  turnover time $\tau^+_{\rm nl}$ (dash) and the nominal eddy turnover
time $\tau^-_{\rm nl}$ (dash-triple-dot). 
  \label{fig:timescales}}
\end{figure*}

\vspace*{0.5cm}
\subsection{Timescales}
\label{sec:ts} 

The $z^+$ AWs injected into the domain at $r=r_{\rm min}$ propagate
away from the Sun and generate $z^-$ waves via non-WKB
reflection. These $z^-$ waves undergo further reflections,
producing more $z^+$ fluctuations, and also interact nonlinearly with
the $z^+$ AWs, producing a turbulent cascade.  In each of our
simulations, we evolve the Els\"asser fields until they reach an
approximate statistical steady state, which occurs after approximately
4~hours of physical time in Simulations A1 through E1, and 5 hours of
physical time in Simulations A2 through~E2.  This approach to steady
state is illustrated in Figure~\ref{fig:Esaturation}, which shows the
time evolution of the total $z^\pm$ energy
\begin{equation}
E_{\rm tot}^\pm = \frac{1}{4}\, \int d^3 x\, \rho |\vec{z}^\pm|^2,
\label{eq:defEpm} 
\end{equation} 
where the volume integral is over the entire simulation domain.

An important timescale for these simulations is the $z^+$ Alfv\'en crossing time
$\tau_{\rm A}$, which is the time it takes a $z^+$ fluctuation to
travel from $r_{\rm min}$ to $r_{\rm max}$, 
\begin{equation}
  \tau_{\rm A} = \int_{r_{\rm min}}^{r_{\rm max}}\frac
  {dr}{U(r)+v_{\rm A}(r)} \simeq 1.3 \mbox{ hours}.
\end{equation}
We define the nominal nonlinear timescale for the large-scale $z^\pm$
fluctuations through the equation 
\begin{equation}
\tau_{\rm nl}^\pm(r) = (k_{\perp \rm min} z^\mp_{\rm rms})^{-1},
\label{eq:deftaunl} 
\end{equation} 
where $k_{\perp \rm min} = 2\pi /L_{\perp}(r)$ is the minimum
perpendiucular wavenumber at radius~$r$. This is the approximate
timescale on which $z^\pm$ fluctuations at perpendicular
scale~$L_\perp$ are sheared by $z^\mp$ fluctuations at perpendicular
scale~$L_{\perp}$, where $z^\mp_{\rm rms}(r)$ is the rms value of
$z^\mp$ at heliocentric distance~$r$.  We define the correlation
time~$\tau_{\rm c}^\pm$ of the (outer-scale) $z^\pm$ fluctuations at
radius~$r$ through the equation
\begin{equation}
C^\pm(r,\tau_{\rm c}^\pm) = e^{-1},
\label{eq:deftauc} 
\end{equation} 
where
\begin{equation}
C^\pm(r,\tau) = \frac{\langle \vec{z}^\pm(\vec x_\perp,r,t)\cdot
  \vec{z}^\pm(\vec x_\perp,r,t+\tau)\rangle }{\langle
  \vec{z}^\pm(\vec x_\perp,r,t)\cdot \vec{z}^\pm(\vec x_\perp,r,t) \rangle},
\label{eq:defC} 
\end{equation} 
 and $\langle \dots \rangle$ denotes an
average over~$x$, $y$, and~$t$.  We also define the reflection
timescale
\begin{equation}
\tau_{\rm r} = \left(\frac{dv_{\rm A}}{dr}\right)^{-1}
\label{eq:deftaur} 
\end{equation} 
that characterizes the $H_{\rm A}^{-1}$ term in
Equation~(\ref{eq:velli2}), which is the only linear term
in Equation~(\ref{eq:velli2})  that couples $\vec{z}^+$ with~$\vec z^-$.

We plot the timescales $\tau_{\rm nl}^\pm$, $\tau_{\rm c}$, and
$\tau_{\rm r}$ as functions of~$r$ in Figure~\ref{fig:timescales}. In
Simulations A1 through C1 and A2 through C2, $\tau_{\rm c\,\sun}^+$
exceeds~$\tau_{\rm r}(R_{\sun})$, which causes reflection to be very
efficient near~$R_{\sun}$. The nonlinear timescale~$\tau_{\rm
  nl}^+$ is significantly larger than~$\tau_{\rm nl}^-$ at all~$r$ and
in all simulations because $z^-_{\rm rms} < z^+_{\rm rms}$. In
Simulations~A1, A2, B1, and~B2, $\tau_{\rm c}^+$ is significantly
larger than $\tau_{\rm nl}^+$ at small~$r$ due to the large coherence
time of the the waves that are launched at $r=R_{\sun}$.

\begin{figure*}[t]
  \centering
  \includegraphics[width=0.475\textwidth]{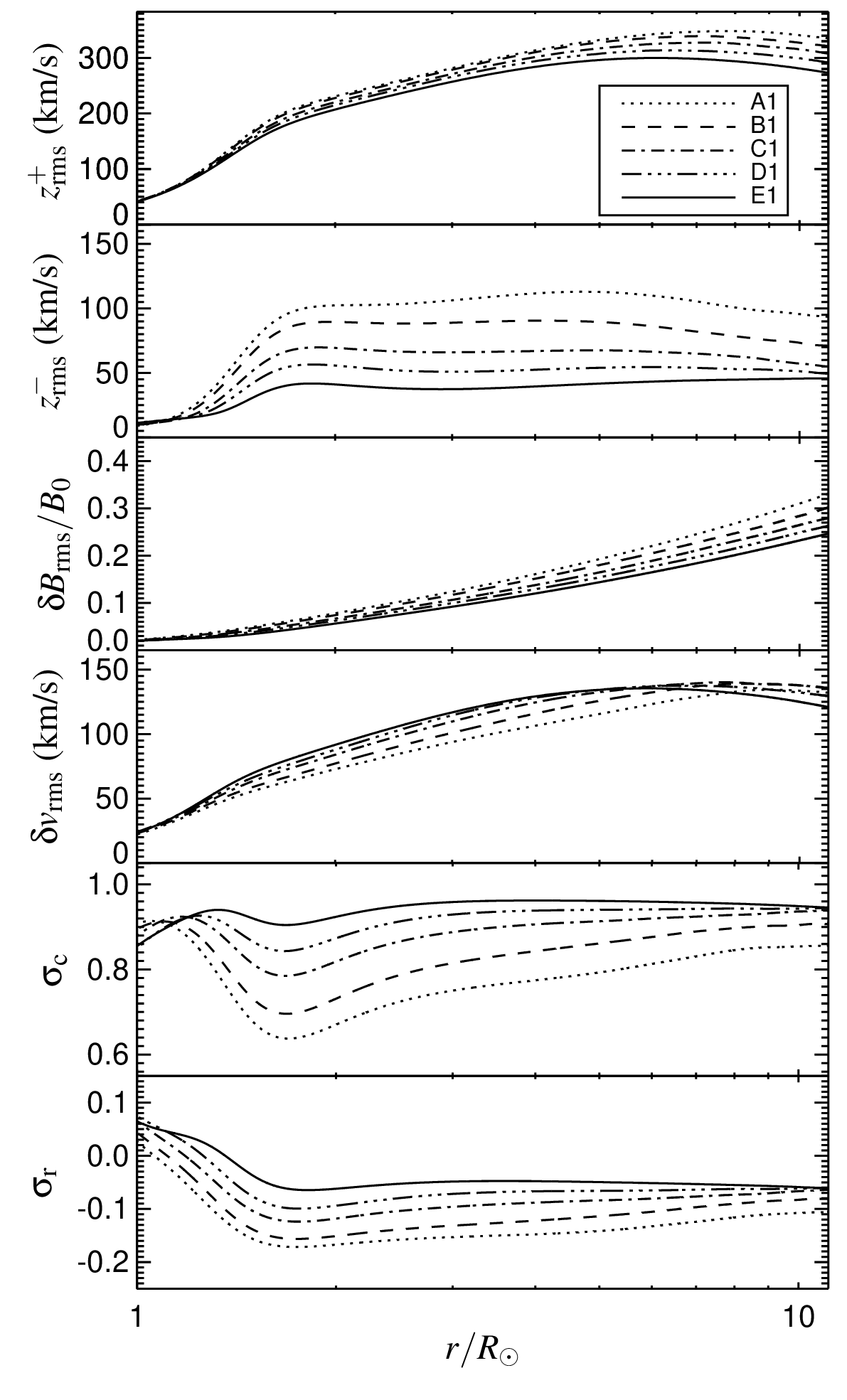}
  \includegraphics[width=0.475\textwidth]{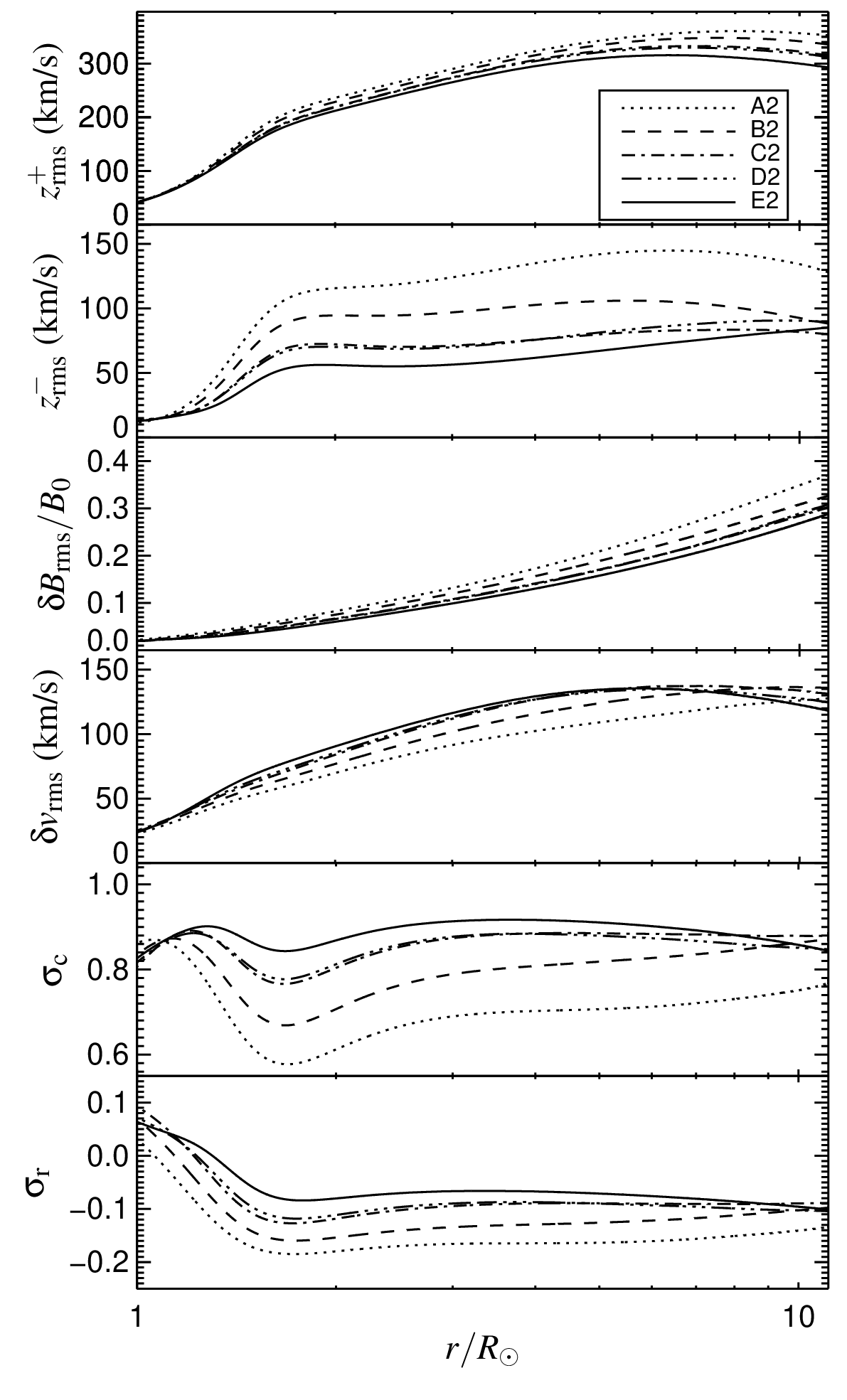}
  \caption{Top four panels show the radial dependence of the rms amplitudes of
    $\vec{z}^\pm$, $\delta \vec{B}$, and $\delta \vec{v}$.
The bottom two panels show the radial profiles of the fractional cross-helicity
    $\sigma_{\rm c}$ and the fractional residual energy $\sigma_{\rm r}$.}
  \label{fig:radial_profiles}
\end{figure*}

The correlation times~$\tau_{\rm c}^+$ and $\tau_{\rm c}^-$ are
comparable to each other at all $r$ exceeding $\sim 2 R_{\sun}$ in
Simulations A1 through D1 and A2 through D2. In these simulations,
$\tau_{\rm nl}^- < \tau_{\rm c}^+$, and the outer-scale $z^+$
fluctuations imprint their correlation time onto the outer-scale $z^-$
fluctuations through nonlinear interactions. In contrast, in
Simulations~E1 and~E2, $\tau_{\rm nl}^- > \tau_{\rm c}^+$ at
large~$r$, and nonlinear interactions are too weak for the outer-scale
$z^+$ fluctuations to transfer their correlation timescale to the
outer-scale $z^-$ fluctuations. In these simulations at large~$r$,
$\tau_{\rm c}^- > \tau_{\rm c}^+$, presumably because radial
inhomogeneities preferentially reflect the lower-frequency component
of~$z^+$.

We note that after the simulations have reached an approximate
statistical steady state, we continue to evolve the fields for an
additional time of at least~$\simeq 7 \mbox{ hours}$. This additional
time is $\gtrsim 5 \tau_{\rm A}$, $\gtrsim 18\tau^+_{\rm c \sun}$ and
$\gtrsim 86\tau_{nl,\sun}^+$ in all simulations. When we take time averages,
we restrict the time averaging to this approximate-steady-state period.

\subsection{Radial Profiles}
\label{sec:radial_profiles}

In Figure~\ref{fig:radial_profiles}, we plot the radial profiles 
of $z^\pm_{\rm rms}$, the rms values of~$\delta \vec{v}$ and $\delta \vec{B}$ (denoted $\delta v_{\rm rms}$ and $\delta B_{\rm rms}$), 
the fractional cross helicity
\begin{equation}
\sigma_{\rm c}(r) = \frac{[z^+_{\rm rms}(r)]^2 -  [z^-_{\rm rms}(r)]^2}{[z^+_{\rm rms}(r)]^2 + [z^-_{\rm rms}(r)]^2},
\label{eq:sigmac}   
\end{equation} 
and the fractional residual energy
\begin{equation}
\sigma_{\rm r}(r) = \frac{4\pi \rho [\delta v_{\rm rms}(r)]^2 -  [\delta B_{\rm rms}(r)]^2}
{4\pi \rho [\delta v_{\rm rms}(r)]^2 +  [\delta B_{\rm rms}(r)]^2}.
\label{eq:sigmar} 
\end{equation} 
The fractional cross helicity is highest for the runs with the
smallest values of $\tau_{\rm c\,\sun}^+$, because shorter correlation
times at the coronal base translate to higher wave frequencies
throughout the domain, which reduces the efficiency of non-WKB wave
reflection.  The maximum value of $\delta B/B_0$ occurs in
Simulation~A1 at $r=r_{\rm max}$, where $\delta B/B_0 = 0.37$. The
assumption that $\delta B \ll B_0$ underlying
Equation~(\ref{eq:velli2}) is thus at least marginally satisfied in
all of the simulations.  The sign of the residual energy depends
upon~$r$. At $r< 1.6 R_{\sun}$, $dv_{\rm A}/dr > 0$ and wave
reflections act to create positive residual energy. At $r> 1.6
R_{\sun}$, $dv_{\rm A}/dr <0$ and wave reflections are a source of
negative residual energy. Because the $z^-$ waves propagate towards
smaller~$r$ after being produced by reflections, the transition from
positive to negative~$\sigma_{\rm r}$ occurs at~$r<1.6 R_{\sun}$.

\vspace*{0.25cm}
\subsection{Power Spectra}
\label{sec:power_spectra} 

Once the fluctuations reach a statistical steady state, the average
$z^\pm$ energy density per unit mass,
\begin{equation}
  u^\pm(r)=\frac 14\aave{\vec z^\pm(\vec x_\perp,r,t)\cdot\vec z^\pm(\vec x_\perp,r,t)},\label{eq:edensity}
\end{equation}
is a function of~$r$ alone. As in section~\ref{sec:ts}, the average is
taken over $x,y$ and $t$. After expanding $\vec{z}^\pm$ in the Fourier series
\begin{equation}
\vec{z}^\pm(\vec{x}_\perp,r,t) = \sum_{\tilde{k}_x}\;\sum_{\tilde{k}_y}
\tilde{\vec{z}}^\pm_{\tilde{k}_x \tilde{k}_y}(r,t)e^{i\vec{k}_\perp \cdot \vec{x}_\perp},
\label{eq:Fseries}
\end{equation} 
where $\tilde{k}_x$ and $\tilde{k}_y$ (defined in Equation~(\ref{eq:defkperpt}))
take on integer values only,
we can rewrite Equation~(\ref{eq:edensity}) as
\begin{eqnarray}
  u^\pm(r)&=&\sum_{\tilde{k}_x} \;\sum_{\tilde{k}_y}
\frac 14\aave{\left|\tilde\vec z^\pm_{\tilde{k}_x \tilde{k}_y}(r,t)\right|^2}\label{eq:edensity2}.
\end{eqnarray}
The average in Equation \eref{eq:edensity2} is now
only over time. 
Because there is no preferred direction in the $x-y$ plane in our
simulations (neglecting anisotropic discretization effects
associated with our Cartesian grid),  the steady-state average 
$\frac 14\aave{\left|\tilde\vec z^\pm_{\tilde{k}_x\tilde{k}_y}(r,t)\right|^2}$ in
Equation~\eref{eq:edensity2} 
is expected to be independent of the direction of~$\vec{k}_\perp$. In
practice, we compute averages only over finite intervals of time, and
thus the right-hand side of Equation~(\ref{eq:edensity2})
does depend to some degree on the wavevector direction. To minimize this
effect and improve our statistics, we compute the 1D power spectrum  
\begin{equation}
  e^\pm(k_\perp,r)=\frac{\pi \tilde k_\perp }{2}\overline{\aave{\left|\tilde\vec
        z^\pm_{\tilde{k}_x \tilde{k}_y}(r,t)\right|^2}},
\end{equation}
where the over-bar denotes the average over the direction of $\vec
k_\perp$.  The normalization
factor $\pi\tilde{k}_\perp/2$ is introduced so that summing over $\tilde{k}_\perp$
 gives $u^\pm(r)$. 
Finally, we define the volume-integrated energy spectrum
$E^\pm$ in the $n^{\rm th}$ radial subdomain (see
Section~\ref{sec:method} for further details) through the equation
\begin{equation} 
  E^\pm(k_\perp,r_n) = \int_{r_n - 0.5\Delta r_n}^{r_n + 0.5\Delta r_n}dr\,\rho a e^\pm,
\label{eq:defEpmn} 
\end{equation} 
where 
\begin{equation}
a(r) = [L_\perp(r)]^2
\label{eq:defa} 
\end{equation} 
is the cross-sectional area of the simulation domain at radius~$r$. As
in Section~\ref{sec:method}, $r_n$ is the radial midpoint of the
$n^{\rm th}$ subdomain, and $\Delta r_n \simeq 0.02 R_{\sun}$ is the
length of the $n^{\rm th}$ subdomain in the $r$ direction.  From the
energy spectrum defined by Equation~\eref{eq:defEpmn}, one can obtain
the total energy in the system by summing over all wavevectors in each
subdomain, and then totaling the energies from all subdomains.

\begin{figure}[!t]
  \centering
  \includegraphics[width=0.5\textwidth]{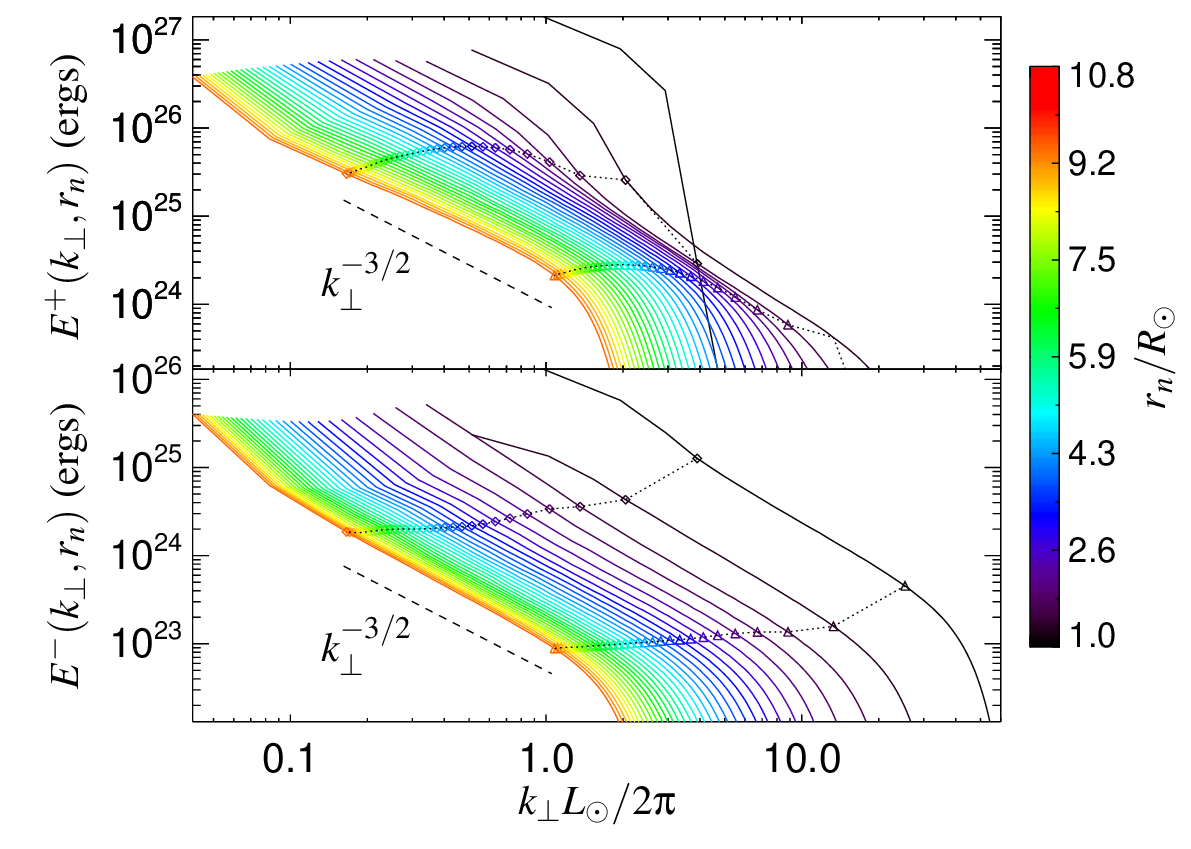}
  \caption{Radial evolution of the steady state energy spectra for simulation E2 in table ~\ref{tab:simlist}.} 
  \label{fig:espectrum_vs_r}
\end{figure}

Figure~\ref{fig:espectrum_vs_r} shows $E^\pm(k_\perp,r_n)$ for
simulation E2 for selected $r_n$ values ranging from $ 1.01 R_{\sun}$ to
$11.09 R_{\sun}$. 
If we were to plot $E^\pm(k_\perp, r_n)$ versus $\tilde
k_\perp=k_\perp L_\perp(r)/2\pi$, the spectra at different $r_n$ would
all cover the same range of $\tilde k_\perp$ values.  However, in
Figure~\ref{fig:espectrum_vs_r} we plot $E^\pm$ versus $k_\perp
L_{\perp \sun}/2\pi$, and thus the spectra migrate to the left as $r$
increases. The spectra extend over the wavenumber range $1 \leq
\tilde{k}_\perp \lesssim 85$ rather than the range $1 \leq
\tilde{k}_\perp \leq 128$, because we set $\Omega^\pm_{ \alpha
  \tilde{k}_x \tilde{k}_y, n} = 0$ when $\tilde{k}_\perp > N_\perp/3$
in order to dealias the simulations, where $\tilde{k}_\perp$ is
defined in Equation~(\ref{eq:defkperpt}). 

\begin{figure*}[!t]
  
  \begin{tabular}{cc}
    \includegraphics[width=0.48\textwidth]{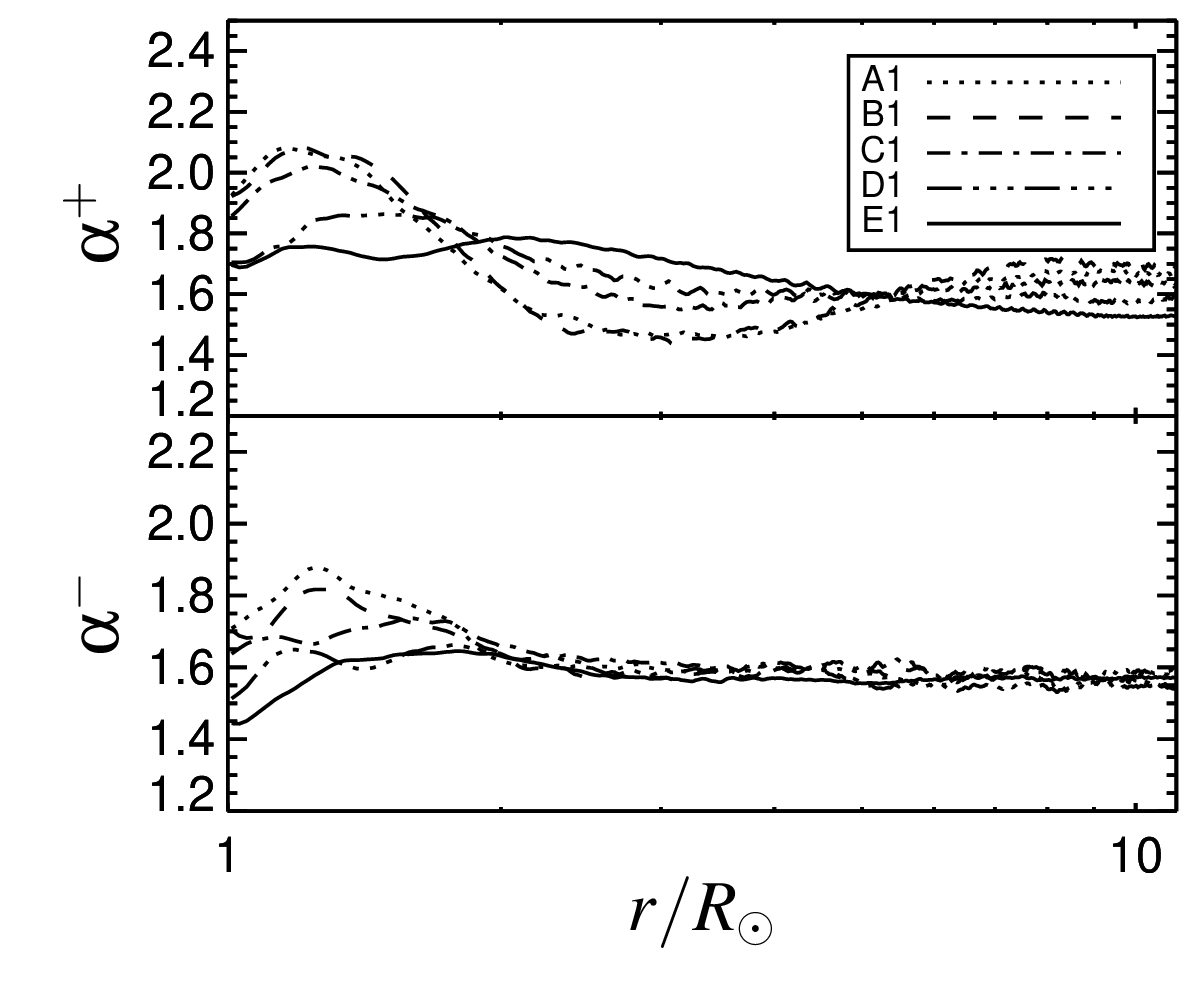}&\includegraphics[width=0.48\textwidth]{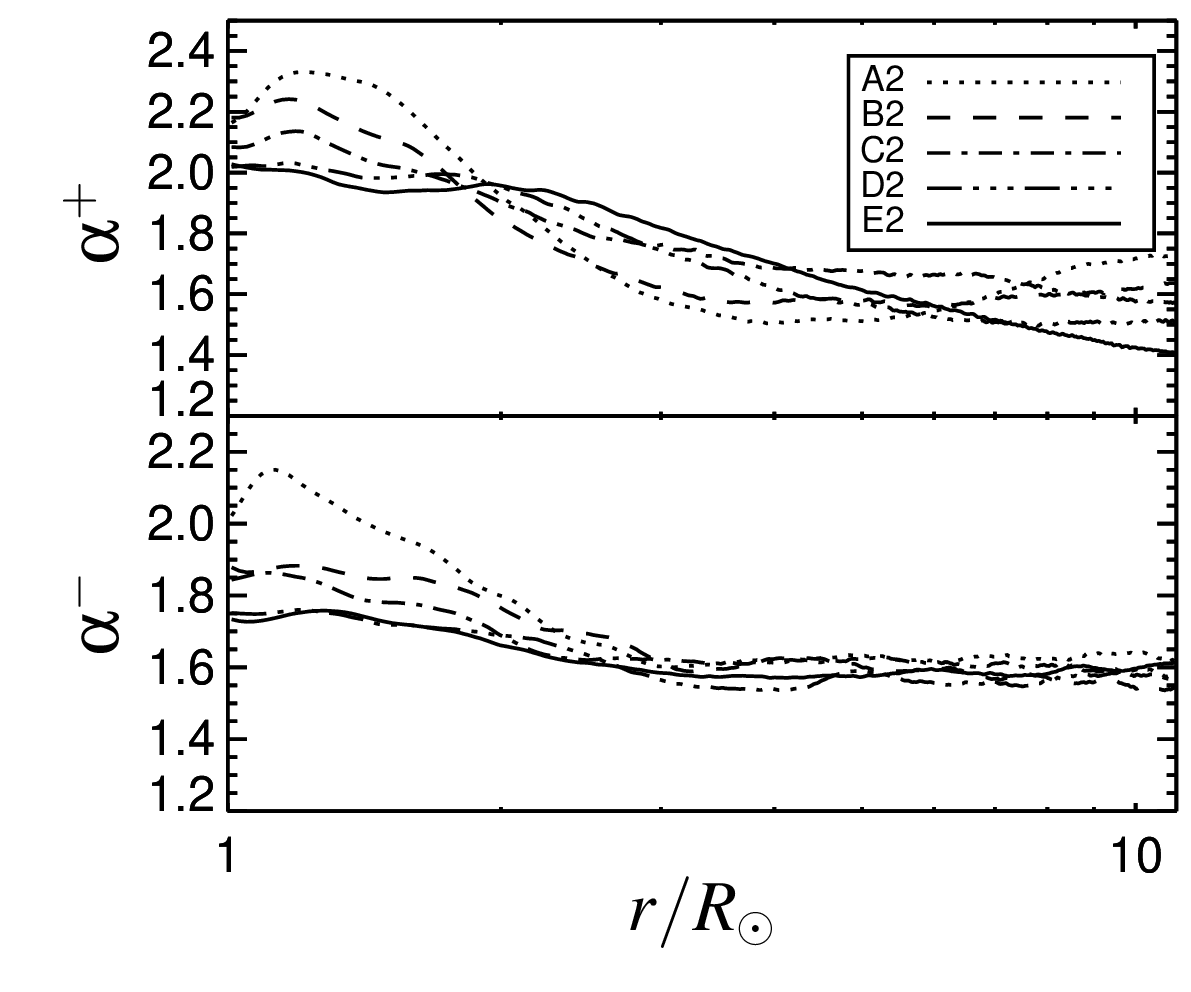} 
  \end{tabular}
  \caption{Radial dependence of the spectral indices~$\alpha^\pm$ of
    the $E^\pm$ energy spectra for
    simulations A1 to E1 (left) and simulations A2 to E2 (right). }\label{fig:slopes_vs_r}
 
\end{figure*}

Because all the energy is injected through the lower boundary and into
wavevectors in the range $1\le \tilde {k}_\perp\le 3$, all
fluctuations on scales $\tilde {k}_\perp \ge4$ are the result of the
nonlinear interaction between $z^+$ fluctuations and the $z^-$
fluctuations that are produced by reflection. As 
Figure~\ref{fig:espectrum_vs_r} illustrates, the fluctuations develop
power-law-like spectra over the wavenumber interval~$4 \lesssim \tilde
{k}_\perp \lesssim 25$ at all locations except  the immediate vicinity
of~$r=R_{\sun}$.  We determine the spectral index~$\alpha_n^\pm$ of
the $z^\pm$ power spectrum in the $n^{\rm th}$ radial subdomain by
fitting $E^\pm(k_\perp,r_n)$ to a power law of the form
\begin{equation}
E^\pm(k_\perp,r_n)\propto k_\perp^{-\alpha_n^\pm}
\label{eq:defalphan} 
\end{equation}
over the wavenumber interval $4 < \tilde{k}_\perp < 25$. Figure
~\ref{fig:slopes_vs_r} shows how $\alpha_n^\pm$ depends upon $r_n$ for
the ten simulations listed in Table~\ref{tab:simlist}.  Although the
large-scale $z^+$ eddies injected at the base of the corona
result in vanishing wave power at $\tilde {k}_\perp > 4$ at $r=r_{\rm
  min}$, there is a non-negligible amount of power at $\tilde{k}_\perp
> 4$ in~$E^\pm(k_\perp, r_1)$ because of the domain averaging
procedure in Equation~(\ref{eq:defEpmn}), which averages the strictly
large-scale spectrum at $r=r_{\rm min}$ with the energy spectrum that
develops at slightly larger radii within the first domain, which
extends out to a maximum radius of $\simeq 1.02 R_\sun$. 
The high-$k_\perp$ $z^+$ fluctuations at small~$r$ arise in part from
the cascade of~$z^+$ fluctuations and in part from the reflection of
broad-spectrum $z^-$ fluctuations at small~$r$~\citep{verdini09a,verdini12}. 
 We discuss the
dependence of~$\alpha^\pm_n$ on $r_n$, $L_{\perp \sun}$, and $\tau_{\rm
  c \sun}^+$ in Section~\ref{sec:anomalous} .

\subsection{Heating Rates and Energy Conservation}
\label{sec:heating_rates}

We obtain a steady-state energy-conservation relation by taking the
dot product of Equation~(\ref{eq:velli2}) with $\rho\vec{z}^\pm$,
adding the resulting equations for $\rho(\partial/\partial
t)(z^+)^2/2$ and $\rho(\partial/\partial t)(z^-)^2 /2$, and dividing
by~2.  We then integrate over the simulation volume and average over
time to obtain
\begin{equation}
F_{\sun}^+ = F_{\rm A}^+ + |F^-_{\sun} |
+ W_{\rm bg} + Q,
\label{eq:econ} 
\end{equation} 
where $F_{\sun}^\pm = F^\pm(R_{\sun})$, $F_{\rm A}^+ =
F^+(r_{\rm A})$, and
\begin{equation} 
F^\pm(r) = \rho a\(U\pm v_{\rm A}\)u^\pm 
\label{eq:defFpmsun} 
\end{equation}
is the area-integrated $z^\pm$ energy flux through
the flux-tube cross section at heliocentric distance~$r$. The quantities
$u^\pm$ and $a$ are
defined in Equations~(\ref{eq:edensity}) and (\ref{eq:defa}), 
$u_{\rm R}\equiv\aave{\vec z^+\cdot\vec z^-}/2$ is the residual energy
density per unit mass, and
\begin{equation}
 W_{\rm bg} = \int dr \rho aU\left(\frac{u^++u^-}{2H_\rho}-\frac{u_{\rm R}}{H_{\rm A}}\right)
\label{eq:defWdot} 
\end{equation} 
is the rate at which the fluctuations do work on the background flow
(e.g. solar-wind acceleration by the magnetic pressure of the AWs). The term
\begin{equation}
   Q = \int_{\rm r_{\rm min}}^{r_{\rm max}} dr \rho a  (q^+ + q^-)
\label{eq:defQpm} 
\end{equation} 
is the total turbulent heating rate integrated over the simulation volume,
\begin{equation}
q^\pm(r) = \frac 12 \nu_p \langle  |{\bf T}^\pm|^2 \rangle
\label{eq:defqpm} 
\end{equation} 
is the $z^\pm$ dissipation power per unit mass at radius~$r$ due to
hyperviscosity, and  
$$
|{\bf T}^\pm|^2 = \sum_{i_1 = 1}^{2} \sum_{i_2=1}^2 \dots \sum_{i_p=1}^2 \sum_{j=1}^2
\big[\(\nabla_\perp\)_{i_1}\cdots\(\nabla_\perp\)_{i_p}z^\pm_j\big]\big[
$$
\begin{equation}
\(\nabla_\perp\)_{i_1}\cdots\(\nabla_\perp\)_{i_p}z^\pm_j\big],
\end{equation}
where $\(\nabla_\perp\)_1 = \partial/\partial x$,
$\(\nabla_\perp\)_2 = \partial/\partial y$, $z^\pm_1 = \ex \cdot \vec{z}^\pm$,
and $z^\pm_2 = \ey \cdot \vec{z}^\pm$.

\begin{figure*}[!t]
\begin{center}
\vspace*{-0.1in}
\includegraphics[width=0.49\textwidth]{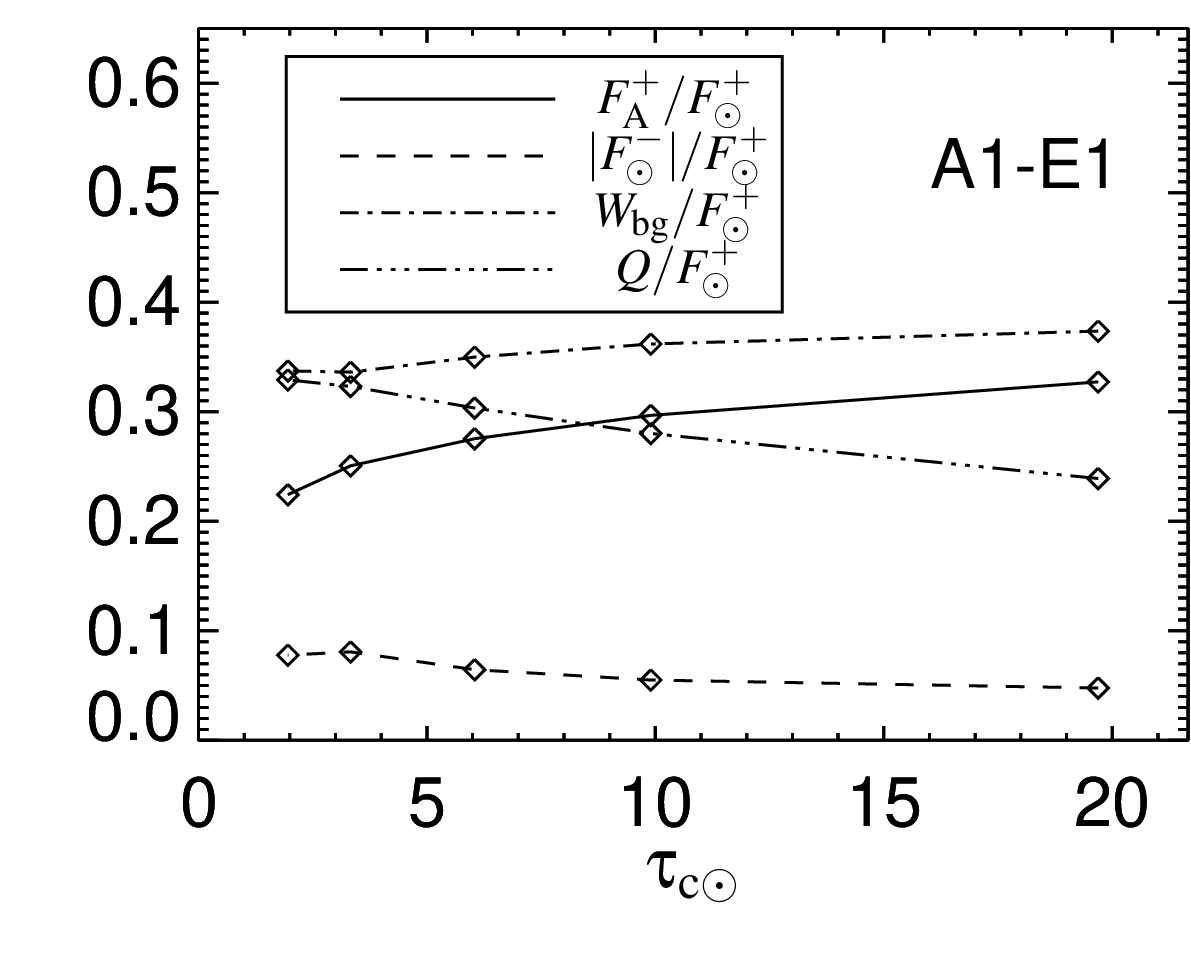}
\includegraphics[width=0.49\textwidth]{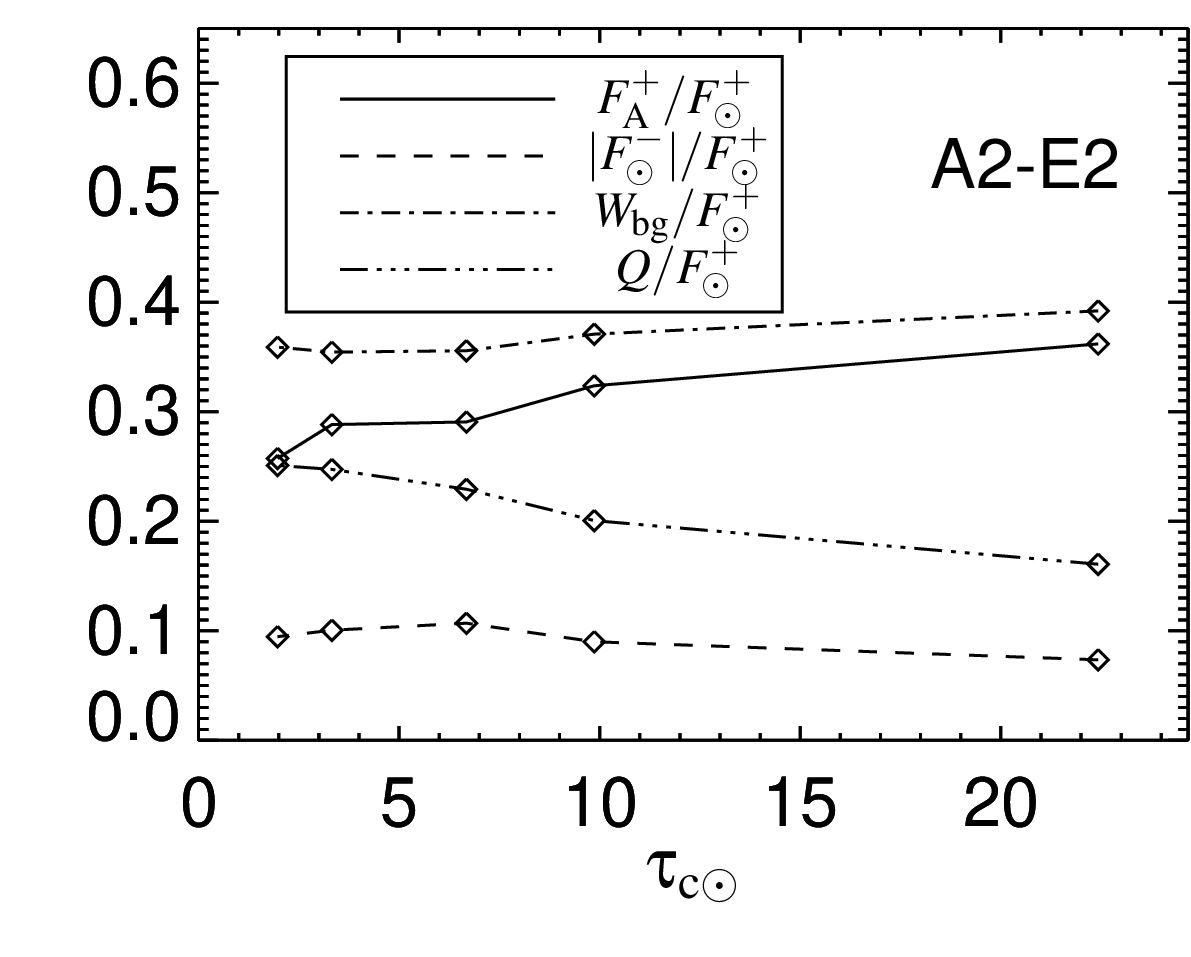}
\end{center}
\caption{\footnotesize The fractions of the $z^+$ input energy at
  $r=R_{\sun}$ that escapes as $z^+$ energy at $r=r_{\rm A}$ (solid
  line), escapes as $z^-$ energy at $r=R_{\sun}$ (dashed line), that
  goes into work on the background flow (dash-dot-dash line), and that
  goes into turbulent heating (dash-triple-dot-dash
  line).
\label{fig:e_cons} }
\end{figure*}

We refer to $F^+_{\sun}$ as the ``input power'' and the time
integral of $F^+_{\sun}$ as the ``input energy.'' In
Figure~\ref{fig:e_cons} we plot the quantities $F^+_{\rm
  A}/F^+_{\sun}$, $|F^-_{\sun}|/F^+_{\sun}$, $W_{\rm bg}/ F^+_{\sun}$, 
and $Q/ F^+_{\sun}$. Equation~(\ref{eq:econ}) implies that in
steady state these fractions add to~one. In our simulations, these
fractions add to $1 \pm 0.03$. The small deviations from unity are
expected because time averages in our finite-duration simulations are
only approximations of a statistical steady state. In Simulations A1
through~E1, the largest single ``sink'' of input energy is the work
done on the background flow, although comparable amounts of energy go
into turbulent heating and the $z^+$ energy that escapes through the
outer boundary at $r=r_{\rm A}$. In Simulations A2 through~E2, the
larger value of~$L_{\perp}$ systematically weakens turbulent
dissipation relative to simulations A1 through E1, and the fraction of
the input power that goes into turbulent heating is reduced to values
$\sim 0.15 - 0.25$.

\begin{figure*}[!t]
\begin{center}
\vspace*{-0.1in}
\includegraphics[width=0.45\textwidth]{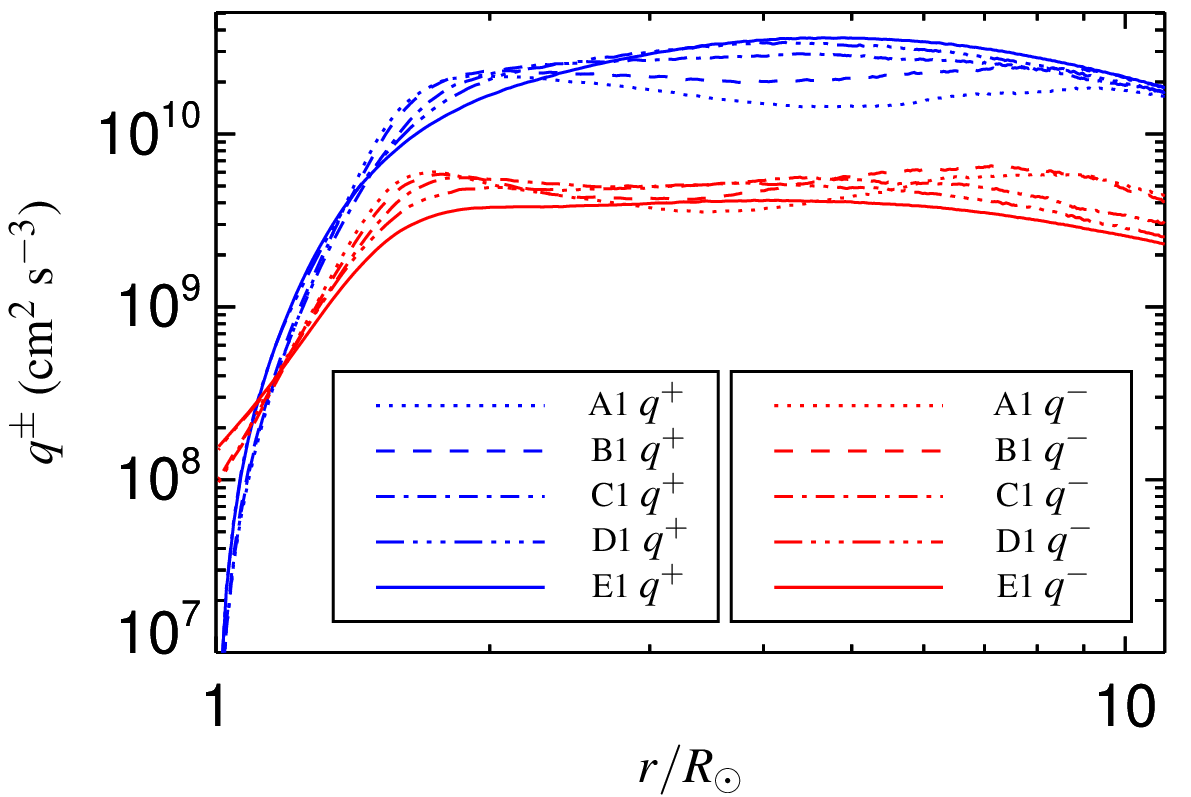}
\includegraphics[width=0.45\textwidth]{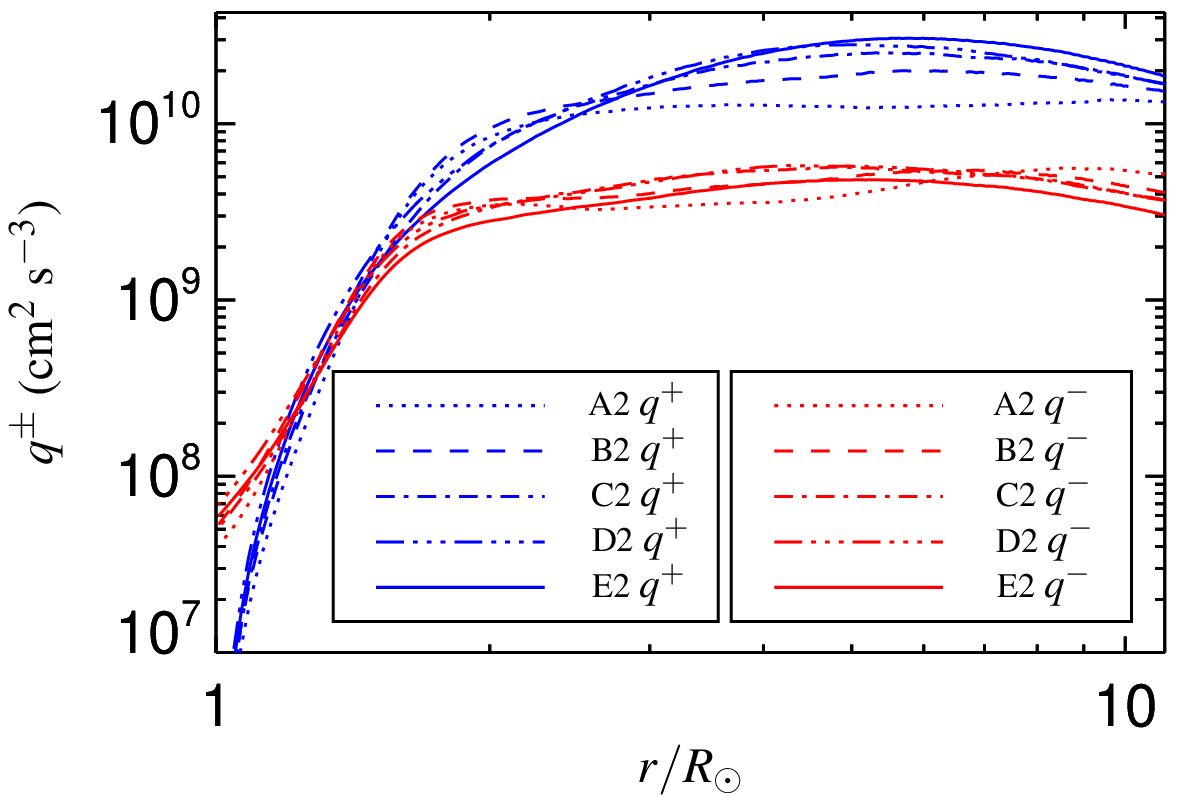}
\end{center}
\caption{\footnotesize Heating-rate profiles~$q^\pm(r)$ (defined in
  Equation~(\ref{eq:defqpm})). {\em Left panel:} Simulations A1
  through~E1, in which $L_{\perp 0 } = 10^4~\mbox{ km}$. {\em Right
    panel:} Simulations A2 through~E2, in which $L_{\perp \sun} =
  2\times 10^4 \mbox{ km}$. 
\label{fig:heating_profiles} }
\end{figure*}

In Figure~\ref{fig:heating_profiles} we plot $q^+(r)$ and $q^-(r)$ for
the simulations in Table~\ref{tab:simlist}.  Even though we have
launched only large-scale AWs at $r=R_{\sun}$, these waves undergo a
vigorous cascade leading to heating rates $\gtrsim 10^{10} \mbox{
  cm}^3 \mbox{ s}^{-1}$ between $2 R_{\sun}$ and $11 R_{\sun}$.  The
reason that $q^+$ decreases as $r$ decreases from $2R_{\sun}$ to
$R_{\sun}$ is that it takes time for the $z^+$ fluctuations entering
the domain at $r=R_{\sun}$ to cascade from the injection
lengthscale~$\sim L_\perp$ to the dissipation scale~$d$, and the $z^+$
fluctuations move away from the Sun during this time. The radius
$r_{\rm q}$ at which $q^+$ reaches its maximum value is larger in the
simulations with $L_{\perp \sun} = 2\times 10^4 \mbox{ km}$ than in
the simulations with $L_{\perp \sun} = 10^4 \mbox{ km}$ because the
energy-cascade timescales increase with increasing $L_{\perp
  \sun}$. The value of $r_{\rm q}$ also depends on the dissipation
scale $d$. In a hypothetical simulation with much higher numerical
resolution and much smaller hyperviscosity, $d$ would be smaller, it
would take longer for energy to cascade from scale~$L_\perp$ to
scale~$d$, and $r_{\rm q}$ would increase. 
On the other hand, in many models of turbulence the energy-cascade
timescale decreases with scale, in which case decreasing~$d$ would
have only a modest effect on the total energy cascade time. If we
were to launch a broad spectrum of $z^+$ AWs into the simulation
domain at $r=r_{\rm min}$, the AWs launched at perpendicular
wavenumbers satisfying $L_\perp^{-1} \ll k_\perp \ll d^{-1}$ would
require less time to cascade to scale~$d$, leading to additional
heating close to the Sun's surface. The radial profile of $q^+$ in
the region $R_{\sun} < r < 2R_{\sun}$ is thus sensitive to the
assumed $k_\perp$ spectrum of the waves launched by the Sun
$e^+(k_\perp, R_{\sun})$. Future investigations into the dependence
of $q^+(r)$ on $e^+(k_\perp, R_{\sun})$ will be important for
determining the radial profile of the heating rate at small~$r$.

We note that $q^+ \gg q^-$ at $r\gtrsim 2 R_{\sun}$ because there is
more $z^+$ energy than~$z^-$ energy. On the other hand, $q^- \gg q+$
very close to $r=r_{\rm min}$, because there is very little $z^+$
energy at large~$k_\perp$ due to the steepness of the $z^+$ power
spectrum near the coronal base.

The peak heating rates of $\sim 10^{10} \mbox{ cm}^2 \mbox{ s}^{-3}$
to $\sim 3 \times 10^{10} \mbox{ cm}^2 \mbox{ s}^{-3}$ in our
simulations are smaller than the peak heating rates of $\sim \times
10^{11} \mbox{ cm}^2 \mbox{ s}^{-3} $ to $\sim \times 10^{12} \mbox{
  cm}^2 \mbox{ s}^{-3}$ in the wave-driven solar-wind models
of~\cite{cranmer07}, \cite{verdini10}, and \cite{chandran11}. One
reason for this difference is that the rms wave amplitudes in our
simulations are smaller than in these models. We do not conclude that
the peak values of~$q^\pm$ in our simulations are more accurate than
in these previous models, because of the uncertainty in the correct
value of $z^+_{\rm rms}$ at the coronal base (see discussion at the
beginning of Section~\ref{sec:simulations}), and to a lesser degree
because of uncertainties in the background solar-wind profiles.

\cite{dmitruk03} conjectured that turbulent heating in coronal holes
at $R_{\sun} < r < 2R_{\sun}$ is favored when the inequalities $
\tau_{\rm nl}^- <\tau_{\rm r} < R_{\sun}/\langle v_{\rm A} \rangle <
\tau_{\rm c\,\sun}^+ < t_{\eta}$ are satisfied, where $t_{\eta}$ is the
resistive timescale at scale~$L_\perp$, and $\langle v_{\rm A} \rangle
$ is the average value of $v_{\rm A}$ between $r=R_{\sun}$ and
$r=2R_{\sun}$. Most of these inequalities are at least marginally
satisfied in our simulations at $r< 2R_{\sun}$. The exception is that
the inequality $R_{\sun}/\langle v_{\rm A} \rangle < \tau_{\rm
  c\,\sun}^+ $ is violated when when $\tau_{\rm c\,\sun}^+ \lesssim 5.84 \mbox{
  min}$. 
The fact that the heating rates at $1.5 R_{\sun} \lesssim r <
2R_{\sun}$ are lower in Simulations D1, D2, E1 and E2 than in the
other simulations is thus consistent with Dmitruk \& Matthaeus' (2003)
conjecture.

The dependence of $q^+$ on~$\tau_{\rm c,\sun}^+$, however, reverses as
$r$ increases to values exceeding~$\sim 2.5 R_{\sun}$. As shown in
Figures~\ref{fig:e_cons} and~\ref{fig:heating_profiles}, runs with
larger~$\tau_{\rm c\sun}^+$ have smaller heating rates at $r\gtrsim
2.5 R_{\sun}$, despite the fact that~$z^-_{\rm rms}$ is larger. In
Section~\ref{sec:vpalign}, we describe how this result can be
understood by considering an effect that systematically weakens 
nonlinear interactions at $r\gtrsim 2.5 R_{\sun}$ as $\tau_{\rm
  c\,\sun}^+$ is increased.

\vspace{0.2cm} 
\section{Anomalous $z^-$ Fluctuations and the $k^{-1}$ Spectrum of the Interplanetary Magnetic Field}
\label{sec:anomalous} 
\vspace{0.2cm} 

One of the most important ideas in the study of reflection-driven
turbulence is the distinction between ``classical'' and ``anomalous''
$z^-$ fluctuations~\citep{velli89,verdini09a,verdini12}.  After a
$z^-$ fluctuation is produced by reflection, it
propagates towards the Sun at speed~$v_{\rm A}$ in the reference frame
of the solar wind. However, because the sources of the $z^-$
fluctuations are $z^+$ fluctuations that propagate away from the Sun
in the solar-wind frame, if one views the pattern of $z^-$
fluctuations, some component of that pattern propagates away from the
Sun along with~$z^+$. This component of the $z^-$ wave field is the
anomalous component.  The remainder of the $z^-$ wave field is the
classical component.

The anomalous $z^-$ fluctuations play an important role because they
coherently shear the $z^+$ fluctuations that produce them.  This
effect is absent in homogeneous RMHD turbulence and alters the
phenomenology of the energy cascade. In homogeneous RMHD turbulence,
nonlinear interactions proceed through a series of collisions between
counter-propagating $z^+$ and $z^-$ wave packets. Successive
collisions are uncorrelated, and so if the wave amplitudes are
sufficiently small, the effects of successive collisions  add
incoherently as in a random walk~\citep{kraichnan65}, leading to a
weak-turbulence regime with a slow energy
cascade~\citep{ng96,ng97,galtier00}.  In contrast, when a $z^+$ wave
packet is sheared by the ``anomalous'' $z^-$ fluctuations produced by
the reflection of that $z^+$ wave packet, the shearing is coherent in
time, which strengthens the nonlinear interaction relative to the
homogeneous case.

\cite{velli89} argued that an energy cascade driven by anomalous
$z^-$ fluctuations leads to an inertial-range $z^+$ power spectrum
that is  much flatter than in homogeneous turbulence. They considered
fluctuations with an isotropic wavenumber spectrum and estimated the
magnitude of the source term for anomalous $z^-$ fluctuations at
wavenumber $k$ to be $z^+_{k}/\tau_{\rm r}$, where $z^+_{k}$ is the
rms amplitude of the $z^+$ fluctuations at perpendicular
scale~$k^{-1}$. They multiplied this source term by
the time it takes for $z^-$ fluctuations to propagate away from their
source, which is $\sim (k v_{\rm A})^{-1}$, to obtain the estimate
$z^-_{k, \rm a} \sim z^+_k/(k v_{\rm A} \tau_{\rm r})$,
where $z^-_{k, \rm a}$ is the rms amplitude of anomalous $z^-$
fluctuations at scale~$k^{-1}$. Upon taking the $z^+$
cascade power
\begin{equation}
\epsilon^+ \sim k z_{k,\rm a}^- (z_k^+)^2
\label{eq:eps1} 
\end{equation} 
to be independent of~$k$, they found that $z_k^+$ becomes independent
of~$k$, leading to a $k^{-1}$ inertial-range power spectrum
for~$z^+$. Since $z^+$ dominates the fluctuation energy, the $k^{-1}$
spectrum for~$z^+$ implies a $k^{-1}$ spectrum for the magnetic field,
velocity field, and fluctuation energy.

The arguments of \cite{velli89} can be revised to account for
wavenumber anisotropy and the shearing of $z^-$ fluctuations by $z^+$
fluctuations. If the fluctuations vary much more rapidly in the
directions perpendicular to~$\vec{B}_0$ than in the direction parallel
to~$\vec{B}_0$ (the ``quasi-2D'' case), then we can replace $k$ with
$k_\perp$ in the arguments of \cite{velli89}, defining $z_{k_\perp}^+$
to be the rms amplitude of $z^+$ fluctuations at perpendicular
scale~$k_\perp^{-1}$, and $z^-_{k_\perp,\rm a}$ to be the rms
amplitude of anomalous $z^-$ fluctuations at perpendicular
scale~$k_\perp$.  The source term for the production of $z^-$ by the
reflection of~$z^+$ at scale~$k_\perp^{-1}$ is $S_{k_\perp} \sim
z^+_{k_\perp}/\tau_{\rm r}.$ The $z^-$ fluctuations at perpendicular
scale~$k_\perp$ cascade to smaller scales in a time~$t_{k_\perp}\sim
(k_\perp z^+_{k_\perp})^{-1}$. If this cascade time is shorter than
the wave period, then $z^-_{k_\perp, \rm a}$ is approximately equal to
the product of $S_{k_\perp}$ and $t_{k_\perp}$; i.e., $z^-_{k_\perp,
  \rm a} \sim (k_\perp \tau_{\rm r})^{-1}$.  Equation~(\ref{eq:eps1})
(with $k$ replaced by~$k_\perp$) then yields the relation
\begin{equation}
\epsilon^+ \sim (z_{k_\perp}^+)^2/\tau_{\rm r}.\label{eq:eps2} 
\end{equation} 
Taking $\epsilon^+$ to be independent of~$k_\perp$ implies that
$z_{k_\perp}^+$ is independent of~$k_{\perp}$. The $z^+$ energy per
unit $k_\perp$, roughly~$(z_{k_\perp}^+)^2/k_\perp$, is then again
$\propto k_{\perp}^{-1}$.

The above arguments suggest that $E^+$ evolves towards a~$k^{-1}$
scaling (or a $k_\perp^{-1}$ scaling in the case of quasi-2D turbulence)
when the anomalous component of $z^-$ dominates
the nonlinear shearing of the $z^+$ fluctuations, and that $E^+$ is
similar to the spectrum of homogeneous RMHD turbulence, in which
$\alpha^+\approx 3/2$, 
when the classical component of $z^-$ dominates the nonlinear shearing
of the~$z^+$ fluctuations\footnote{A number of numerical simulations
  of homogeneous RMHD turbulence and homogeneous MHD turbulence have
  found $\alpha^+ \simeq
  3/2$~\citep{maron_g01,haugen_04,muller_g05,mason_cb06,mason_cb08,perez_b10_2,chen11,perez12},
  but others have reported values closer to
  $5/3$~\citep{beresnyak_l06,chen_11,beresnyak_12}. This discrepancy
  has been discussed in some detail by Perez et al (2012).}.  To
explain the behavior of the power   
spectra in our simulations, we would need to explain the relative
contributions of anomalous and classical $z^-$ fluctuations to the
shearing of~$z^+$ at each scale, which is beyond the scope of this
paper. We instead confine ourselves to the following observations. 

\begin{figure*} 
  \centering
  \begin{tabular}{cc}
  \includegraphics[width=0.5\textwidth]{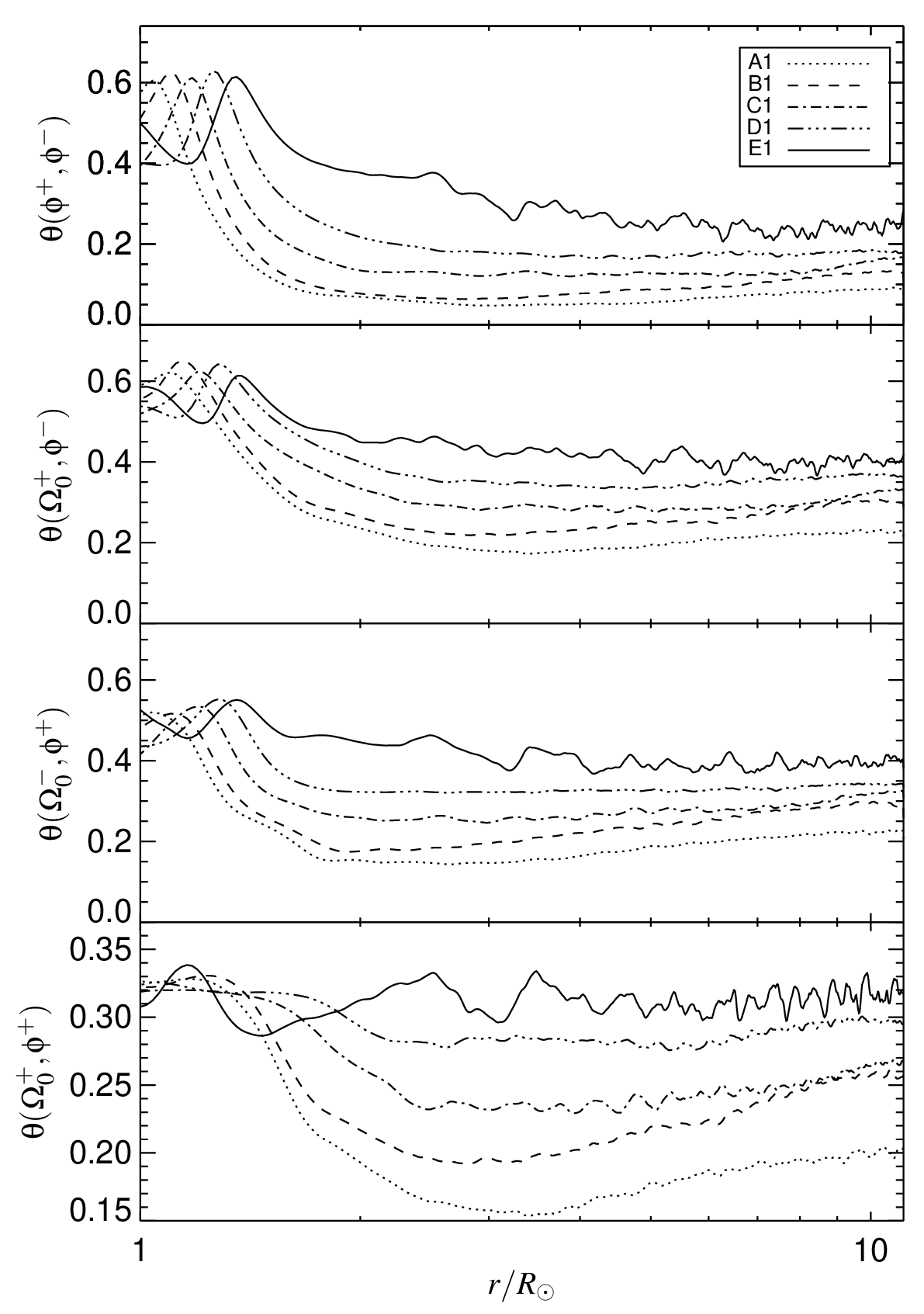}&
  \includegraphics[width=0.5\textwidth]{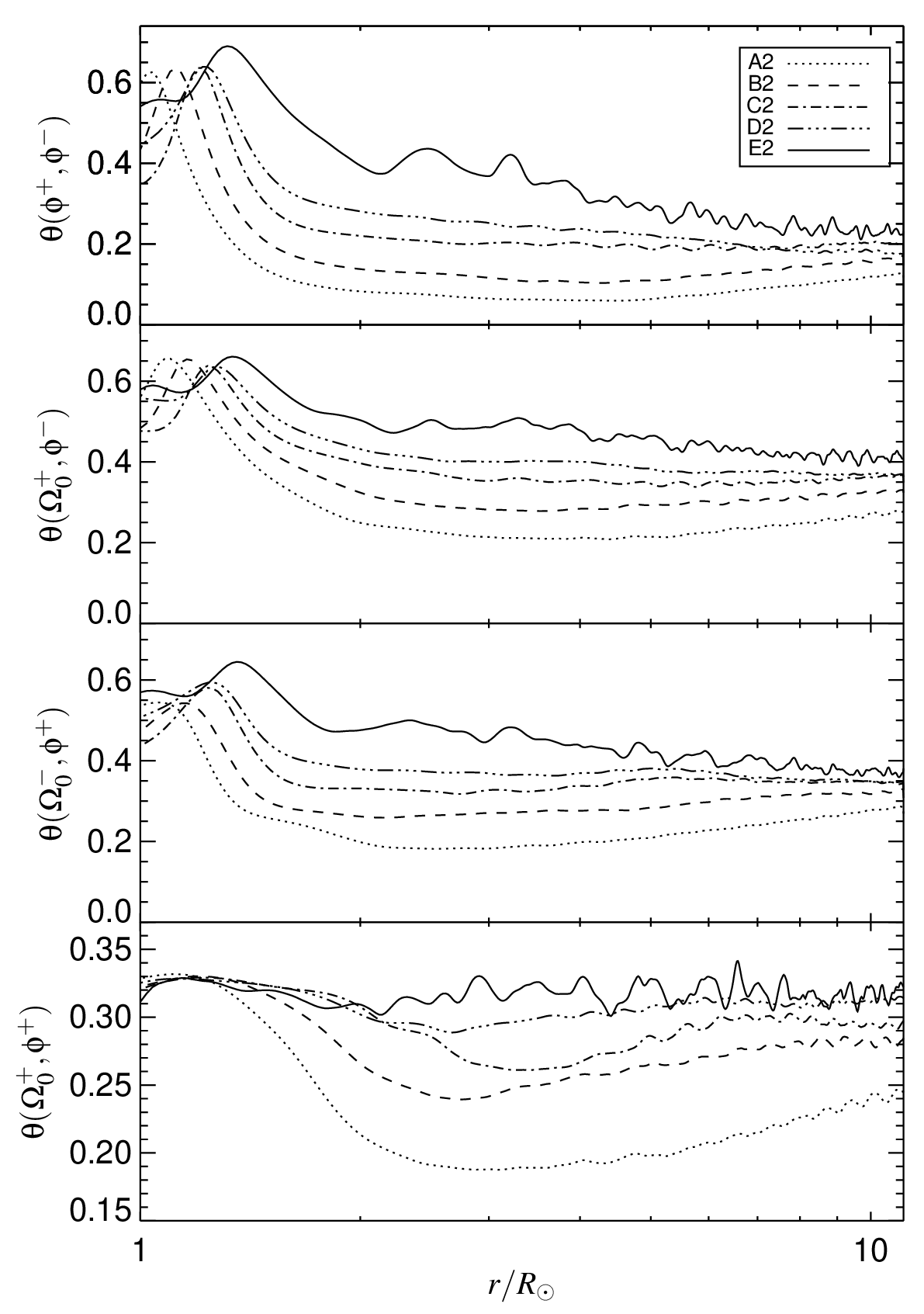}
  \end{tabular}
  \caption{Alignment angles for the Els\"asser potentials~$\phi^\pm$ and
outer-scale vorticity~$\Omega^\pm_0$, which excludes contributions
to~$\Omega^\pm$ from $\tilde{k}_\perp > 2$. 
The left panel corresponds to simulations A1 through E1, and the right panel to Simulations A2 through~E2. 
\label{fig:VPangles}}
\end{figure*}

The flat spectra in our simulations arise in two distinct regimes.
First, in the large-$\tau_{\rm \,\sun}^+$ runs A1, A2, and~B1,
$\alpha^+$ decreases to values of $1.3 - 1.4$ at $3R_{\sun} < r < 4
R_{\sun}$.  These flat spectra are in some ways reminiscent of the
$z^+$ spectrum found by \cite{verdini09a} in numerical simulations of
reflection-driven AW turbulence in which the nonlinear terms in
Equation~(\ref{eq:velli2}) were approximated using a shell model. In
their simulations, $\alpha^+ \sim 1.2$ at $r\lesssim 1.2 R_{\sun}$,
but at larger radii the $z^+$ spectrum steepened towards $\alpha^+ =
5/3$. Second, in our smallest-$\tau_{\rm c\,\sun}^+$ simulations (E1
and~E2), the spectrum becomes increasingly flat as~$r$ increases. The
clearest evolution towards a $k_\perp^{-1}$ spectrum occurs in
Simulation~E2, in which $\alpha^+$ decreases almost linearly
with~$\ln(r)$, reaching the value $\alpha^+=1.3$ at $r =
11.1R_{\sun}$. The upper-right-hand panel of
Figure~\ref{fig:slopes_vs_r} suggests that $\alpha^+$ would reach
even smaller values in Simulation~E2 if we were to extend that
simulation to larger~$r$.

A characteristic that sets Simulations E1 and~E2 apart from the other
simulations is that the $z^-$ cascade becomes weak at large~$r$. This
can be seen from the bottom panels of Figure~\ref{fig:timescales},
which show that $\tau_{\rm c}^+ < \tau_{\rm nl}^-$ at large~$r$ in
simulations E1 and~E2.  This inequality means that outer-scale $z^+$
fluctuations decorrelate in less time than the time it takes them to
significantly distort the outer-scale~$z^-$ fluctuations.  As a
consequence, the effects on~$z^-$ of successive collisions between
outer-scale wave packets accumulate in a random-walk-like manner,
which slows down the $z^-$ energy cascade.  Another characteristic
that distinguishes Simulations~E1 and~E2 is the lack of alignment
between contours of constant~$\phi^+$ and contours of constant
(outer-scale)~$\Omega^+$, as illustrated in the bottom panel of
Figure~\ref{fig:VPangles}.  As we discuss further in
Section~\ref{sec:vpalign}, the alignment between these contours that
arises in the other simulations acts to weaken the nonlinear
interactions arising from anomalous~$z^-$ fluctuations.

The flat spectra in Simulations~E1 and~E2 may be relevant to the
magnetic power spectrum~$E_B(k)$ observed in the interplanetary
medium.  Spacecraft measurements show that the power spectrum at small
wavenumbers (i.e., neglecting the dissipation range) has a
broken-power-law form, with $E_B(k) \propto k^{-n_1}$ at $k < k_{\rm
  br}$ and $E_B(k) \sim k^{-n_2}$ at $k > k_{\rm br}$, where $n_1
\simeq 1$ and $n_2 \simeq 5/3$ \citep{matthaeus86,tumarsch95}.
Moreover, $k_{\rm br}$ increases as $r$ decreases from 1~AU to
0.3~AU~\citep{bruno05}. The results of Simulations E1 and~E2 support
the suggestion of \cite{velli89} that reflection-driven turbulence can
give rise to a $k^{-1}$ scaling of~$E_B$ over at least some ranges
of~$k$ and~$r$.  Our results, however, are not fully conclusive on
this point because the wavenumber spectrum of $z^+$ at large~$r$
depends upon the frequency spectrum of~$z^+$ at the coronal base,
which is uncertain.  \cite{cranmer05} analyzed time series of the
observed positions of magnetic bright points (MBPs) on the Sun and
found that MBP motions had dominant timescales of $1 - 10 \mbox{
  min}$. The timescales of photospheric motions, however, may be
larger than the timescale~$\tau_{\rm c\,\sun}^+$ characterizing
outer-scale fluctuations at the coronal base. Van Ballegooijen et
al~(2011) \nocite{vanballegooijen11} carried out numerical simulations
of AW turbulence from the photosphere to the corona. They found that
when large-scale waves are launched from the photosphere, these waves
become fully turbulent within the chromosphere, before reaching the
transition region. This causes the outer-scale fluctuations in their
simulations to vary on a timescale that is significantly shorter in
the corona than at the photosphere (see their Figure~8c). On the other
hand, fluctuations in the Faraday rotation of radio signals passing
near the Sun suggest that the outer-scale magnetic fluctuations at
$2R_{\sun} \lesssim r \lesssim 15 R_{\sun}$ vary on timescales
of~$\sim 1 \mbox{ hr}$~\citep{hollweg82}.  Pinning down the frequency
spectrum of the waves at the coronal base and incorporating
broad-spectrum AW launching into numerical simulations such as the
ones presented here will be important for clarifying the possible
connection between reflection-driven turbulence and the $k^{-1}$ power
spectra observed in the solar wind.

\vspace{0.2cm}
\section{Vorticity-Potential Alignment}
\label{sec:vpalign} 
\vspace{0.2cm} 

The nonlinear term on the right-hand side of
Equation~\eref{eq:vorticity}, $ \mathcal N^\pm \equiv 
-\eb\cdot\left[\nabla_\perp\times\(\vec
z^\mp\cdot\nabla_\perp\vec z^\pm\)\right]$, 
is responsible for the transfer
of energy from large scales to small scales. This term can be
written in the form~\citep{schekochihin09}
\begin{equation} 
\mathcal N^\pm=\frac 12\(\[\Omega^+,\phi^-\]+\[\Omega^-,\phi^+\]\pm\nabla_\perp^2\[\phi^+,\phi^-\]\),
\label{eq:defN} 
\end{equation} 
where 
\begin{equation}
\[f,g\]\equiv\ez\cdot\(\nabla_\perp f\times\nabla_\perp g\)
\end{equation}
is the Poisson bracket for arbitrary scalar functions $f$ and~$g$.
These equations show that the shearing of $\vec z^\pm$ by $\vec
z^\mp$ is related to the angle between $\nabla_\perp \phi^+$ and
$\nabla_\perp \phi^-$, the angle between $\nabla_\perp \Omega^+$ and
$\nabla_\perp\phi^-$, and the angle between $\nabla_\perp \Omega^-$
and $\nabla _\perp \phi^+$. We define the characteristic alignment
angle~$\theta(f,g)$ for the scalar functions $f$ and $g$ through the
equation
\begin{eqnarray}
  \theta(f,g) = \sin^{-1}\left(\frac{\aave{\left|\nabla_\perp f\times\nabla_\perp g\right|}}{\aave{\left|\nabla_\perp f\right|\left|\nabla_\perp g\right|}}\right),
\label{eq:defangles} 
\end{eqnarray}
where $\langle \dots \rangle$ denotes an average over $x$, $y$, and~$t$.
From equation~\eref{eq:defN} it follows that the angles
$\theta(\phi^\pm, \Omega^\mp)$ and $\theta(\phi^+,\phi^-)$ are
intimately related to the strength of the
interaction between $z^\pm$ fluctuations, therefore it is natural to
expect that any decrease in these angles results in a weakening of
nonlinear interactions. 

\begin{figure*} 
  \centering
  \includegraphics[width=0.95\textwidth]{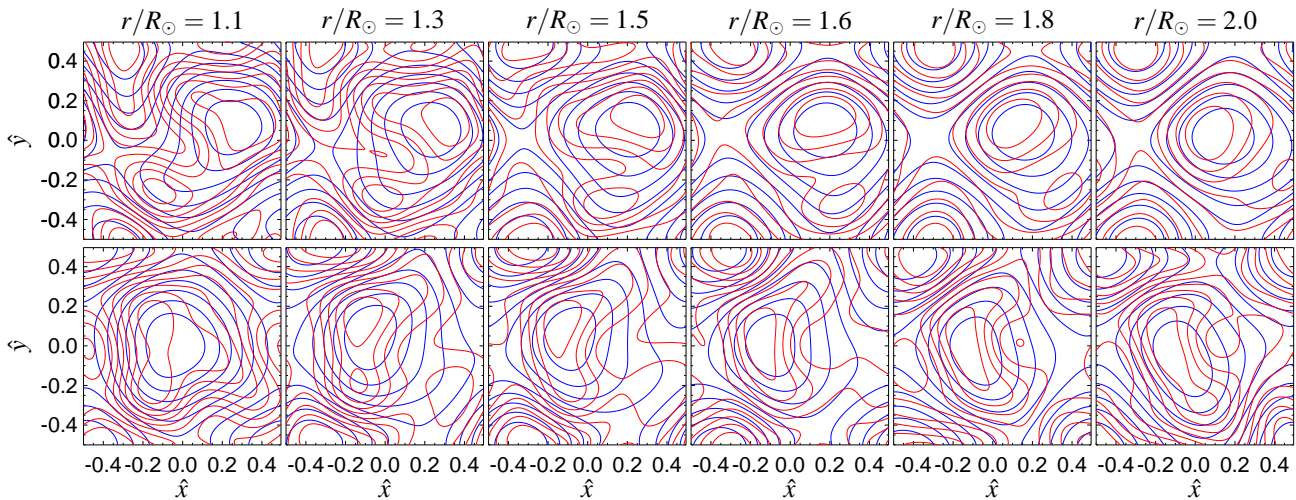}
  \caption{Contours of constant $\phi^+$ (blue line) and constant $\Omega_0^+$ (red line) in field-perpendicular planes at selected radii ($r/R_\sun=1.1,1.3,1.5,1.6,1.8,2.0$). The top panels correspond to simulation A1 and the bottom panels corresponds to simulation D1. The contours of $\phi^+$ and $\Omega_0^+$ clearly become more aligned at larger values of $\tau_{\rm c\sun}$. In these plots $\hat x=x/L_\perp(r)$ and $\hat y=y/L_\perp(r)$.
\label{fig:VPanglecontours}}
\end{figure*}

Figure \ref{fig:VPangles} shows the alignment angles
$\theta(\phi^+,\phi^-)$, $\theta(\Omega_0^+, \phi^-)$,
$\theta(\Omega_0^-, \phi+)$, and $\theta(\Omega_0^+, \phi^+)$ as
functions of~$r$ for the runs in Table~\ref{tab:simlist}, where
$\Omega_0^\pm$ is the value of~$\Omega^\pm$ after we have filtered out
the contributions from~$\tilde{k}_\perp > 2$.
This filtering process allows us to focus on the alignment angles that
characterize nonlinear interactions between outer-scale
fluctuations. In the absence of filtering, $\Omega^\pm$ is dominated
by the largest~$k_\perp$ values in the inertial range, which we assume
have little effect upon the energy cascade at scales~$\sim L_\perp$.

As $r$ increases from $R_{\sun}$ to $3 R_{\sun}$, these angles
decrease, to a degree that increases as
$\tau_{\rm c\,\sun}$ increases. The alignment angles are $\sim 2-4$ times
smaller in Simulation~A1 than in Simulation~E1. On the other hand,
$z^-_{\rm rms}$ is $2-3$ times smaller in Simulation~E1 than in
Simulation~A1.  Vorticity-potential alignment thus helps to explain
why $q^+$ is a decreasing function of~$\tau_{\rm c\,\sun}^+$ at $3
R_{\sun} \lesssim r \lesssim 10 R_{\sun}$, despite the fact that
$z^-_{\rm rms}$ is an increasing function of $\tau_{\rm c\,\sun}^+$.

The dependence of alignment on $\tau_{\rm c\,\sun}^+$ derives in part
from the linear physics of wave reflection. The source term in
Equation~(\ref{eq:vorticity}) representing the production of
$\Omega^-$ by wave reflection has the form $- (U + v_{\rm
  A})\Omega^+/2H_{\rm A}$. The structure of this term in the $x-y$
plane is identical to the structure of $\Omega^+$, and thus reflection
acts to produce an $\Omega^-$ field that is locally ``aligned''
with~$\Omega^+$, in the sense that $\theta(\Omega^+,\Omega^-)$ is
small. For the same reason, reflection acts to produce a $\phi^-$
field that is locally aligned with $\phi^+$, and an~$\Omega_0^-$ field
that is aligned with~$\Omega_0^+$.  After $z^-$ fluctuations are
produced by wave reflection, they propagate towards the Sun at
speed~$v_{\rm A}$ in the plasma frame, separating from the $z^+$
fluctuation that produced them. This separation acts to decrease the
alignment between $\phi^+$ and $\phi^-$ and becomes more important
as~$\tau_{\rm c\,\sun}$ decreases, since the radial lengthscales of
the fluctuations become shorter, enabling $z^-$ fluctuations to
separate from their sources more rapidly.

The bottom panel of Figure~\ref{fig:VPangles} shows the alignment
angle between $\phi^+$ and $\Omega_0^+$. This angle does not
characterize any of the terms in Equation~(\ref{eq:defN}), and so is
not a direct measure of the efficiency of nonlinear
interactions. Instead, the decrease in $\theta(\phi^+, \Omega_0^+)$
with increasing~$r$ in Simulations A1 through D1 and A2 through D2 can
be characterized as the decay of~$z^+$ fluctuations towards a state of
``vorticity-potential alignment.''  As the fluctuations approach a
state in which $\theta(\phi^+,\Omega_0^+)=0$, the outer-scale $z^-$
fluctuations produced by wave reflection become increasingly
inefficient at shearing outer-scale $z^+$ fluctuations and are in turn
sheared increasingly inefficiently by the outer-scale $z^+$
fluctuations. The radial decay of $\theta(\phi^+,\Omega_0^+)$ weakens
as $\tau_{\rm c\,\sun}^+$ decreases.  One reason for this is likely
that $z^-_{\rm rms}$ also decreases with decreasing $\tau_{\rm
  c\,\sun}^+$, so that the ``selective decay'' of the unaligned
component of $z^+$ proceeds more slowly.  The dependence of
$\theta(\phi^+,\Omega_0^+)$ on~$r$ and~$\tau_{\rm c,\sun}^+$ is
illustrated in Figure~\ref{fig:VPanglecontours}, which plots contours
of~$\phi^+$ and~$\Omega_0^+$ at selected radii between~$1 R_{\sun}$
and~$2 R_{\sun}$ in Simulations~A1 and~D1.

\vspace{0.2cm} 
\section{Summary and Conclusion}
\label{sec:conclusion} 
\vspace{0.2cm}

We have carried out the first direct numerical simulations of
inhomogeneous RMHD turbulence from the coronal base to the Alfv\'en
critical point that take into account radial variations in~$U$,
$\rho$, and~$B_0$ without approximating the nonlinear terms in the
inhomogeneous RMHD equations.  The
simulation domain is a magnetic flux tube with a square cross section
of area~$L_\perp(r)^2$ that extends from $r=R_{\sun}$ to $r= r_{\rm
  A}=11.1 R_{\sun}$, the location of the Alfv\'en critical point in
our model solar wind. This flux tube is narrow, with $L_\perp \ll r$
at all~$r$.

There are three control parameters in the simulations: the rms
amplitude of $z^+$ at the coronal base, denoted~$z^+_{\rm rms\,\sun}$,
the correlation time of the outer-scale $z^+$ fluctuations at the
coronal base, denoted $\tau_{\rm c\,\sun}^+$, and the width of the
cross section of the simulation domain at the coronal base,
denoted~$L_{\perp \sun}$.  For the ten simulations reported in this
study, we choose~$z^+_{\rm rms\,\sun}$ so that the rms velocity at
$r=R_{\sun}$ is~$\simeq 20 \mbox{ km/s}$. We take $L_{\perp \sun}$ to
be either $10^4 \mbox{ km}$ or~$2\times 10^4 \mbox{ km}$, comparable
to the spatial scales of supergranules. We consider $\tau_{\rm
  c\,\sun}^+$ values ranging from $2 \mbox{ min}$ to $20 \mbox{ min}$.
(A smaller value of~$\tau_{\rm c\,\sun}^+$ implies higher wave
frequencies, which reduces the efficiency of wave
reflection.)

The~$z^+$ AWs that we launch through the simulation boundary at
$r=R_{\sun}$ have small perpendicular wavenumbers ~$\sim 2\pi/L_{\perp
  \,\sun}$. As the $z^+$ fluctuations propagate away from the Sun,
they undergo partial non-WKB reflection, which generates $z^-$
fluctuations.  Nonlinear interactions between $z^+$ fluctuations and
$z^-$ fluctuations causes fluctuation energy to cascade to small
scales and dissipate, heating the ambient plasma. Between 15\% and
33\% of the $z^+$ energy launched by the Sun (the ``input energy'')
dissipates within the simulation domain, between 33\% and 40\% goes
into work on the background flow, and between 22\% and 36\% escapes as
$z^+$ energy at $r=r_{\rm A}$, for all input parameters investigated
in this work. Our finding that $\sim 1/6 - 1/3$ of the input energy
dissipates at $r< r_{\rm A}$ is consistent with Chandran \&
Hollweg's~(2009)\nocite{chandran09c} analytical model of
reflection-driven turbulence in the solar wind, which finds (their
Equation~43) that the outer-scale $z^+$ fluctuations experience an
order-unity number of cascade times as they propagate from the coronal
base to the Alfv\'en critical point, without any fine-tuning of
parameters.  These results provide an important consistency check on
models in which the solar wind is powered by AWs and AW turbulence. If
only a tiny fraction of the input energy were dissipated at $r< r_{\rm
  A}$, then AW turbulence would be unable to explain the powerful
heating rates that are inferred from observations of ion temperature
profiles in coronal holes~\citep{kohl98,esser99}. If almost all of the
input energy were dissipated, then the wave amplitudes and heating
rates near the Sun would have to be unrealistically large in order
that enough energy would survive to explain the large amplitudes of
outward-propagating $z^+$ fluctuations observed at $r\sim 0.3 \mbox{
  AU}$.

As the $z^+$ fluctuations propagate away from the Sun, their power
spectrum gradually flattens towards power a power law of the
form~$E^+(k_\perp) \propto k_\perp^{-\alpha^+}$, where $1.3 \lesssim
\alpha^+ \lesssim 2.4$. \cite{velli89} argued that reflection-driven
turbulence  gives rise to a $k^{-1}$ power spectrum for the $z^+$
fluctuations, which could potentially explain the $k^{-1}$ magnetic
power spectrum observed at large scales in the solar wind. In our
Simulation~E2, $\alpha^+$ decreases steadily with increasing~$r$,
reaching a value of~$1.3$ at $r=11.1 R_{\sun}$. The steady decline of
$\alpha^+$ with increasing $r$ suggests that the spectrum would
become even flatter if we were to extend the simulation to larger~$r$.
In Simulation~E2, $L_{\perp \sun} = 2\times 10^4 \mbox{ km}$ and
$\tau_{\rm c\,\sun}^+ = 2 \mbox{ min}$. 
Similar behavior is seen in the $z^+$ spectrum in Simulation~E1 in
which $L_{\perp \sun} = 10^4 \mbox{ km}$ and $\tau_{\rm c\,\sun}^+ = 2
\mbox{ min}$, 
but the spectral flattening is less pronounced.  In our other
simulations with larger values of~$\tau_{\rm c\,\sun}^+$, the spectra
at large~$r$ are steeper. The ability of reflection-driven turbulence
to explain the $k^{-1}$ spectra in the solar wind thus depends upon
the frequency and wavenumber spectra of the $z^+$ AWs launched by the
Sun, as well as upon the evolution of the fluctuations at $r > r_{\rm
  A}$.

In the simulations that we have run with $\tau_{\rm c\,\sun}^+ \geq 3.3
\mbox{ min}$,  
the fluctuations develop a type of alignment between the contours of
constant~$\phi^+$, $\phi^-$, $\Omega_0^+$, and $\Omega_0^-$, where
$\Omega_0^\pm$ is the contribution to~$\Omega^\pm$ from the
outer-scale fluctuations, and $\phi^\pm$ and $\Omega^\pm$ are the
Els\"asser potential and Els\"asser vorticity defined in
Equations~(\ref{eq:potential}) and (\ref{eq:vorticity-potential}).  In
Simulations A1 through~D1 and A2 through~D2, these angles decrease
between $r=R_{\sun}$ and $r\sim 2 R_{\sun}$, which causes nonlinear
interactions to weaken.  This effect becomes increasingly pronounced
as~$\tau_{\rm c\,\sun}^+$ increases, and helps to explain why the
turbulent heating rate at $r\gtrsim 3R_{\sun}$ decreases with
increasing~$\tau_{\rm c\,\sun}^+$ despite the fact that $z^-_{\rm
  rms}$ increases, as shown in Figure~\ref{fig:heating_profiles}.

Our simulations with~$L_{\perp \sun} = 10^4 \mbox{ km}$ are broadly
similar to our simulations with~$L_{\perp \sun} = 2\times 10^4 \mbox{
  km}$. Since the perpendicular correlation length of the turbulence
in our simulations is~$\simeq L_\perp$, our simulations show that
modest changes in this perpendicular correlation length lead to only
moderate changes in the properties of reflection-driven turbulence
between the Sun and the Alfv\'en critical point. On the other hand, it
is possible that values of~$L_{\perp \sun}$ much smaller than~$10^4
\mbox{ km}$ would lead to significantly different results, and future
simulations to investigate this possibility would be useful.

Another useful direction for future research would be to carry out
simulations in which a broad spectrum of waves is launched
at~$r=R_{\sun}$. More work is also needed to clarify the phenomenology
of reflection-driven turbulence, to explain the different types of
power spectra that it produces, and to determine whether it provides a
viable explanation for at least some portion of the $k^{-1}$ magnetic
power spectrum observed at small~$k$ in the solar wind. Finally, it
will be informative to compare the results of simulations such as the
ones presented here with future measurements from the Solar Probe Plus
spacecraft, which has a planned perihelion of $9.5 R_{\sun}$ that lies
inside the region that we are simulating numerically. Such comparisons
will be useful for testing theories of reflection-driven AW turbulence
and clarifying the role played by AW turbulence in the origin of the
solar wind.

\acknowledgments We thank A. Schekochihin for discussions of weak MHD
turbulence, M. Velli for discussions of anomalous $z^-$ fluctuations,
and A. van Ballegooijen for discussions of alignment effects in
reflection-driven turbulence.  We also thank the anonymous referee for
a very useful report, which helped us to improve the manuscript. This
work was supported in part by grant NNX11AJ37G from NASA's
Heliophysics Theory Program, NASA grant NNN06AA01C to the Solar Probe
Plus FIELDS Experiment, NASA grant NNX12AB27G, NSF grant AGS-0851005, NSF/DOE grant
AGS-1003451, and DOE grant
DE-FG02-07-ER46372. B.~Chandran was supported in part by a Visiting
Research Fellowship from Merton College, University of
Oxford. High-performance-computing resources were provided 
by the Argonne Leadership Computing Facility (ALCF) at Argonne
National Laboratory, which is supported by the Office of Science of
the U.S. Department of Energy under contract DE-AC02-06CH11357. The
ALCF resources were granted under two INCITE projects in 2012 and
2013. High-performance computing resources were also provided by the
National Institute for Computational Sciences (NICS) at the University
of Tennessee under the NSF-XSEDE Project TG-ATM100031.


\end{document}